\newcommand*{\QUANTUM}{}%
\ifdefined\QUANTUM
\documentclass[a4paper,onecolumn,11pt,accepted=2022-04-11]{quantumarticle}
\pdfoutput=1
\else
\documentclass[11pt]{article}
\fi

\usepackage{geometry}
\usepackage{amsmath}
\usepackage{graphicx}
\usepackage{dcolumn}
\usepackage{epstopdf}
\usepackage{bm}
\usepackage{subfig}
\usepackage{braket}
\usepackage{algorithm}
\usepackage{algorithmic}
\usepackage[sort,square,numbers]{natbib}
\usepackage{adjustbox}
\usepackage{cellspace} 
\usepackage{makecell} 
\setcellgapes{3pt}
\usepackage{qcircuit}
\usepackage{multirow}

\usepackage[T1]{fontenc}
\usepackage{color}
\usepackage{amsthm}
\usepackage{amssymb}
\usepackage{esint}
\usepackage{comment}
\usepackage{hyperref}
\usepackage[capitalize]{cleveref}

\newtheorem{theorem}{Theorem}
\newtheorem{lemma}{Lemma}

\newtheorem{cor}{Corollary}

\newtheorem{definition}{Definition}

\title{Time-dependent Hamiltonian Simulation of Highly Oscillatory Dynamics and Superconvergence for Schr\"odinger Equation}

\author[1]{Dong An}
\author[2,3,5]{Di Fang} 
\author[2,4,5]{Lin Lin}
\affil[1]{Joint Center for Quantum Information and Computer Science (QuICS), University of Maryland, College Park, MD 20742, USA}
\affil[2]{Department of Mathematics, University of California, Berkeley,  CA 94720, USA}
\affil[3]{Simons Institute for the Theory of Computing, University of California, Berkeley, CA 94720, USA}
\affil[4]{Computational Research Division, Lawrence Berkeley National Laboratory, Berkeley, CA 94720, USA}
\affil[5]{Challenge Institute for Quantum Computation, University of California, Berkeley, CA 94720, USA}

\begin{document}

\newcommand{\diag}{\operatorname{diag}}
\renewcommand{\Re}{\operatorname{Re}}
\renewcommand{\Im}{\operatorname{Im}}
\newcommand{\conj}[1]{\overline{#1}}
\newcommand{\Tr}{\operatorname{Tr}}
\newcommand{\Op}{\operatorname{op}}

\newcommand{\I}{\mathrm{i}}

\newcommand{\mc}[1]{\mathcal{#1}}
\newcommand{\mf}[1]{\mathfrak{#1}}
\newcommand{\wt}[1]{\widetilde{#1}}

\newcommand{\abs}[1]{\left\lvert#1\right\rvert}
\newcommand{\norm}[1]{\left\lVert#1\right\rVert}

\newcommand{\ud}{\,\mathrm{d}}
\newcommand{\ad}{\operatorname{ad}}

\renewcommand{\arraystretch}{1.5}

\newcommand{\Or}{\mathcal{O}}
\newcommand{\EE}{\mathbb{E}}
\newcommand{\NN}{\mathbb{N}}
\newcommand{\RR}{\mathbb{R}}
\newcommand{\CC}{\mathbb{C}}
\newcommand{\ZZ}{\mathbb{Z}}

\maketitle

\begin{abstract}

We propose a simple quantum algorithm for simulating highly oscillatory quantum dynamics, which does not require complicated quantum control logic for handling time-ordering operators. To our knowledge, this is the first quantum algorithm that is both insensitive to the rapid changes of the time-dependent Hamiltonian and exhibits commutator scaling. 
Our method can be used for efficient Hamiltonian simulation in the interaction picture. 
In particular, we demonstrate that for the simulation of the Schr\"odinger equation, our method exhibits superconvergence and achieves a surprising second order convergence rate, of which the proof rests on a careful application of pseudo-differential calculus.
Numerical results verify the effectiveness and the superconvergence property of our method.
\end{abstract}

\newpage
\tableofcontents 

\section{Introduction}

Hamiltonian simulation is of immense importance in characterizing the dynamics for a diverse range of systems in quantum physics, chemistry, and materials science. 
It is also used as a subroutine in many other quantum algorithms. 
Let $H(t) \in  \mathbb{C}^{2^{n_s}\times 2^{n_s}}$ be a time-dependent Hamiltonian on the time interval $[0,T]$, where $n_s$ denotes the number of system qubits. 
The problem of Hamiltonian simulation is to solve the time-dependent Schr\"odinger equation 
\begin{equation}\label{eqn:ham_sim}
    \I \partial_t \ket{\psi(t)} = H(t) \ket{\psi(t)}, \quad \ket{\psi(0)} = \ket{\psi_0}. 
\end{equation}
The exact evolution operator is given by $U(t,0) = \mathcal{T} \exp\left(-\I \int_0^t H(s) ds\right)$ where $\mathcal{T}$ is the time-ordering operator. 
When $H(t) \equiv H$ does not depend on time, this is a time-independent Hamiltonian simulation problem, and $U(t,0)=\exp\left(-\I H t\right)$.
If the time evolution operator $U(t,0)$ varies slowly with respect to $t$ (for both time-independent and time-dependent Hamiltonian simulations), many numerical integrators can yield satisfactory performance. 
On the other hand, if $U(t,0)$ is highly oscillatory with respect to $t$, the numerical integrator must be carefully chosen to reduce the computational cost. The high oscillation of $U(t,0)$ has two main sources: (1) the Hamiltonian $H(t)$ itself oscillates rapidly in time, \emph{i.e.}, the spectral norm of the time derivative, $\|H'(t)\|$, is large.  (2) $H(t)$ has high energy modes, i.e. the spectral norm $\|H(t)\|$  is large. This can result in highly oscillatory wavefunctions even if $H$ itself is time-independent. 

A wide range of applications falls into one or both categories.
For example, highly oscillatory wave functions or Hamiltonian with large spectral norm are commonly observed in adiabatic quantum computation~\cite{FarhiGoldstoneGutmannEtAl2000,AlbashLidar2018}, $k$-local Hamiltonians (with a large number of sites)~\cite{LowChuang2017,LowChuang2019,ChildsSuTranEtAl2020}, and the electronic structure problem (with a real space discretization)~\cite{KivlichanWiebeBabbushEtAl2017,KivlichanMcCleanWiebeEtAl2018,AnFangLin2021,SuBerryWiebeEtAl2021}. 
On the other hand, in quantum control problems with ultrafast lasers~\cite{MizrahiNeyenhuisJohnsonEtAl2013,NielsenDowlingGuEtAl2006,DongPetersen2010}, and interaction picture Hamiltonian simulation~\cite{LowWiebe2019,RajputRoggeroWiebe2021}, the Hamiltonian itself typically contains highly oscillatory component. 
We remark that under certain assumptions, a Hamiltonian of large spectral norm can be more effectively simulated in terms of a fast oscillatory Hamiltonian in the interaction picture, which will be further discussed later. 

There have been remarkable progresses in the recent years on designing new algorithms as well as establishing improved theoretical complexity estimate of existing algorithms for time-independent Hamiltonian simulation~\cite{BerryAhokasCleveEtAl2007,BerryChilds2012,BerryCleveGharibian2014,BerryChildsCleveEtAl2014,BerryChildsCleveEtAl2015,BerryChildsKothari2015,LowChuang2017,ChildsMaslovNamEtAl2018,LowWiebe2019,ChildsOstranderSu2019,Campbell2019,Low2019,ChildsSu2019,ChildsSuTranEtAl2020,ChenHuangKuengEtAl2020,SahinogluSomma2020}. 
However, for time-dependent Hamiltonian simulation, there are considerably fewer quantum algorithms available, including Monte Carlo method~\cite{PoulinQarrySommaEtAl2011}, truncated Dyson series method~\cite{BerryChildsCleveEtAl2015,KieferovaSchererBerry2019,LowWiebe2019}, permutation expansion based Dyson series methods~\cite{ChenKalevHen2021}, continuous qDRIFT~\cite{BerryChildsSuEtAl2020}, and rescaled Dyson series methods~\cite{BerryChildsSuEtAl2020} \footnote{When the Hamiltonian can be split into the sum of several simpler Hamiltonians, standard and generalized Trotter methods~\cite{HuyghebaertDeRaedt1990,WeckerHastingsWiebeEtAl2015,WiebeBerryHoyerEtAl2010,AnFangLin2021,PoulinQarrySommaEtAl2011} can also be applied. For now we only focus on the simulation for general time-dependent Hamiltonian $H(t)$, and we postpone the discussion of Trotter type methods later when we discuss the Hamiltonian simulation in the interaction picture.}. 
Among these algorithms, the high order truncated Dyson series method achieves so far the best asymptotic complexity for general time-dependent Hamiltonian simulation as well as unbounded Hamiltonian simulation in the interaction picture~\cite{LowWiebe2019}.
Despite the advantages in terms of asymptotic  scaling, 
the implementation of truncated Dyson series method beyond the first order expansion requires the explicit monitoring of the time ordering operations of the form $\int_0^t d t_1 \int_0^t d t_2 \cdots \int_0^t d t_k \mathcal{T} \left[H(t_1) H(t_2) \cdots H(t_k) \right]$. This leads to complicated quantum control logic~\cite{LowWiebe2019}, and significant overheads as well as large constant factors~\cite{SuBerryWiebeEtAl2021,RajputRoggeroWiebe2021}. Such operations are also incompatible with efficient simulation methods such as qubitization~\cite{LowChuang2019} and quantum singular value transformation (QSVT)~\cite{GilyenArunachalamWiebe2019}. 
These drawbacks significantly hinder the practical efficiency of high order  truncated Dyson series methods, and inspire the recent development of hybridized interaction picture Hamiltonian simulation methods that do not rely on complicated quantum control logic~\cite{RajputRoggeroWiebe2021}.
Our work follows the same trajectory and focuses on practical methods that do not require complicated clock construction circuits.
To the extent of our knowledge, all methods within this category~\cite{BerryChildsSuEtAl2020,RajputRoggeroWiebe2021} achieve first order accuracy.

\vspace{1em}
\noindent\textbf{Contribution:}

In this work, we propose a simple quantum algorithm, called \emph{quantum Highly Oscillatory Protocol} (qHOP).
The derivation of qHOP can be succinctly summarized as follows  (see \cref{sec:qhop_derivation} for more details). On each short time interval $[(j+1)h,jh]$ ($h$ is the time step), we simply replace the time-ordered matrix exponential with the standard matrix exponential. This is a time-independent Hamiltonian simulation problem and can be performed with QSVT. 
We approximate the integral $\int_{jh}^{(j+1)h}H(s)ds$ by a numerical quadrature with a large number of quadrature points $M$, which can depend polynomially on $\|H'(t)\|$. 
We remark that the generous use of quadrature points here is distinctly different from that in any classical integrators used in practice, of which the number of quadrature points must be judiciously chosen to reduce the computational cost~\cite{HochbruckLubich2003,Thalhammer2006,BlanesCasasThalhammer2017}. 
The quantum computer can leverage the efficiency of the linear combination of unitaries (LCU) technique~\cite{ChildsWiebe2012,GilyenSuLowEtAl2019}, and the additional cost only scales logarithmically with respect to $M$.

The effectiveness of qHOP for highly oscillatory dynamics can be intuitively understood as follows. 
First, qHOP can be derived by truncating the Magnus expansion~\cite{Magnus1954} up to the first order. 
Thanks to the efficiency of the LCU technique, we can choose a relatively large number of quadrature nodes with a logarithmic overhead, and the dominant part of the qHOP approximation error is due to that of the truncated Magnus expansion, which can be expressed in terms of the integral of the time-dependent Hamiltonian or its commutators and is independent of the derivatives of the Hamiltonian.
As a result, even when $\|H'\|$ becomes very large, as long as the numerical quadrature is performed sufficiently accurately, the cost of qHOP scales almost linearly with respect to the norm of the commutator $\sup_{s,t}\|[H(s),H(t)]\|$, and is independent of  $\|H'\|$.
In many applications, the norm of the commutator can be smaller than $\sup_t \|H(t)\|^2$~\cite{ChildsSuTranEtAl2020}. 

Second, when $\|H'\|$ is bounded and $\|H\|$ is large, we prove that qHOP can in fact achieve second order accuracy. This leads to better scalings than first order truncated Dyson series method in almost all parameters, namely the norm $\|H\|$, the evolution time $T$ and the simulation error $\epsilon$. 
This is because the local time discretization error of qHOP explores local commutator scaling $\max_{|s-t|\leq h} \|[H(s),H(t)]\|$ for small time step size $h$, and this can offer an extra order of $h$ when the time derivative $\norm{H'}$ is bounded. 

Third, we show that qHOP can also achieve the $L^1$-norm scaling if the time step sizes can be dynamically chosen in the same fashion as that of the continuous qDRIFT method~\cite{BerryChildsSuEtAl2020}. 
More specifically, we derive an alternative complexity upper bound which scales almost quadratically in the average performance of $\|H(s)\|$, namely $T^{-1}\int_0^T \|H(s)\|ds$. 
Such an $L^1$-norm scaling demonstrates that qHOP can be efficient for certain fast oscillatory Hamiltonians, whose spectral norm can be large at certain time $t$ but is relatively small on average. Therefore the performance of qHOP is at least comparable to that of continuous qDRIFT. Our numerical results  indicate that qHOP can significantly outperform the continuous qDRIFT method in practice.

To the extent of our knowledge, qHOP is the first quantum algorithm that simultaneously exhibits commutator scaling and is insensitive to fast oscillations of $H(t)$. 
\cref{tab:main_result_1} compares qHOP with the Monte Carlo method, the first order truncated Dyson series method and the continuous qDRIFT method. 
The results demonstrate that the scaling of qHOP is better than the existing three algorithms in both scenarios. 

\begin{table}[]
    \centering
    \begin{tabular}{p{2.2cm}|m{5.1cm}|m{3cm}|m{3cm}}\hline\hline
        \multirow{2}{8em}{Methods} & \multicolumn{3}{c}{Query complexities}\\
        \cline{2-4} & General & w. bounded $\|H'\|$ and large $\|H\|$ & w. large $\|H'\|$ \\\hline
        Monte Carlo method & $\widetilde{\Or}\left( \min\left\{\frac{\alpha^2\widetilde{\alpha}^2 T^4}{\epsilon^3}, \frac{\alpha^2\widetilde{\beta}^{1/2} T^{7/2}}{\epsilon^{5/2}}\right\}\right)$ & $\widetilde{\Or}\left( \frac{\alpha^{5/2} T^{7/2}}{\epsilon^{5/2}}\right)$ & $\mathcal{O}\left(\frac{\alpha^2\widetilde{\alpha}^2 T^4}{\epsilon^3}\right)$ \\\hline 
        First order truncated Dyson series & $\Or\left(\frac{\alpha^2 T^2}{\epsilon}\right)$ & ${\Or}\left(\frac{\alpha^2 T^2}{\epsilon}\right)$ & $\Or\left(\frac{\alpha^2 T^2}{\epsilon}\right)$ \\\hline
        continuous qDRIFT & ${\Or}\left(\frac{\overline{\alpha}^2 T^2}{\epsilon}\right) $ & ${\Or}\left(\frac{\overline{\alpha}^2 T^2}{\epsilon}\right)$ & ${\Or}\left(\frac{\overline{\alpha}^2 T^2}{\epsilon}\right)$ \\\hline
        qHOP & $\widetilde{\Or}\left( \min\left\{\frac{\widetilde{\alpha}^2 T^2}{\epsilon}, {\frac{\overline{\alpha}^2 T^2}{\epsilon}},\frac{\widetilde{\beta}^{1/2} T^{3/2}}{\epsilon^{1/2}}\right\}\right) $ $+ \widetilde{\Or}\left(\alpha T\right)$ & $\widetilde{\Or}\left( \frac{\alpha^{1/2} T^{3/2}}{\epsilon^{1/2}} + \alpha T \right) $ & $\widetilde{\Or}\left( \frac{T^2 \min\left\{\widetilde{\alpha}^2,\overline{\alpha}^2\right\}}{\epsilon}\right) + \widetilde{\Or}\left(\alpha T \right) $ \\\hline\hline 
    \end{tabular}
    \caption{Comparison of query complexities of using different methods to simulate \cref{eqn:ham_sim} on the time interval $[0,T]$ within $\mathcal{O}(\epsilon)$ error. Here we assume $\max_{s\in[0,T]} \|H(s)\| \leq \alpha$, $T^{-1}\int_0^T \|H(s)\| ds \leq \overline{\alpha}$, $\max_{s,t\in[0,T]}\|[H(s),H(t)]\| \leq \widetilde{\alpha}^2$, and $\max_{s,t\in[0,T]}\|[H'(s),H(t)]\| \leq \widetilde{\beta}$. Query complexities are measured by numbers of queries to the input model of the time-dependent Hamiltonian. For continuous qDRIFT, the query complexity is measured with respect to the oracle $e^{-iH(t)/p(t)}$ and the error is measured in the diamond norm of quantum channels, instead of the operator norm of unitaries.}
    \label{tab:main_result_1}
\end{table}

\begin{table}[]
    \centering
    \begin{tabular}{p{4cm}|m{6.2cm}|m{3cm}}\hline\hline
        \multirow{2}{8em}{Methods} & \multicolumn{2}{c}{Query complexities}\\
        \cline{2-3} & General & Schr\"odinger  \\\hline
        First order Trotter & $\Or\left(\frac{\|[A,B]\| T^2}{\epsilon}\right)$ & $\Or\left(\frac{ N T^2}{\epsilon}\right)$\\\hline 
        Second order Trotter & $\Or\left(\frac{(\|[A,[A,B]]\|+\|[B,[B,A]]\|)^{1/2} T^{3/2}}{\epsilon^{1/2}}\right)$ & $\Or\left(\frac{ N T^{3/2}}{\epsilon^{1/2}}\right)$ \\\hline 
        {Monde Carlo method (interaction picture)} & {$\mathcal{O}\left(\min\left\{\frac{\alpha_B^4 T^4}{\epsilon^3}, \frac{\alpha_B^{5/2}(\alpha_{AB}+\beta_B)^{1/2}T^{7/2}}{\epsilon^{5/2}}\right\}\right)$}  & {$\mathcal{O}\left(\frac{T^{7/2}}{\epsilon^{5/2}}\right)$} \\\hline 
        First order truncated Dyson series (interaction picture) & {$\Or\left(\frac{\alpha_B^2 T^2\log((\alpha_{AB}+\beta_B)/\alpha_B)}{\epsilon}\right)$} & ${\Or}\left(\frac{ T^2\log(N)}{\epsilon}\right)$ \\\hline 
        Continuous qDRIFT (interaction picture)& $\Or\left(\frac{\alpha_B^2 T^2}{\epsilon}\right)$ & $\Or\left(\frac{ T^2}{\epsilon}\right)$ \\\hline 
        qHOP & $\widetilde{\Or}\Big( \min\Big\{\frac{\alpha_B^2 T^{2}\log(\alpha_{AB}+\beta_B)}{\epsilon},$  & $\widetilde{\Or}\left(\frac{T^{3/2}\log(N)}{\epsilon^{1/2}}\right)$ \\
         (interaction picture)& $\quad\quad\quad\alpha_B T + \frac{\alpha_B^{1/2} (\alpha_{AB}+\beta_B)^{1/2} T^{3/2}} {\epsilon^{1/2}}   \Big\}  \Big)$ &
        \\\hline\hline
    \end{tabular}
    \caption{Comparison of query complexities of using different methods to simulate \cref{eqn:ham_sim_unbounded} on the time interval $[0,T]$ within $\mathcal{O}(\epsilon)$ error. The column ``General'' refers to the scenario where $A$ and $B(t)$ are two arbitrary Hamiltonians such that $\|A\| \leq \alpha_A$ is  large but $e^{-\I A t}$ can be fast-forwarded, $\max_{t\in[0,T]}\|B(t)\| \leq \alpha_B$, $\max_{t\in[0,T]}\|B'(t)\| \leq \beta_B$, and $\max_{t\in[0,T]}\|[A,B(t)]\| \leq \alpha_{AB}$. For simplicity, in first order Trotter and second order Trotter methods, we further assume $B(t) \equiv B$ is time-independent. The column ``Schr\"odinger'' refers to the digital simulation of the Schr\"odinger equation where the Hamiltonian $H$ is discretized from the operator $-\Delta + V(x)$ for a smooth bounded potential function $V(x)$, where $\Delta$ is the Laplacian operator. Specifically, $A$ corresponds to the discretization of $-\Delta$, and $B$ corresponds to the potential. Here $N$ denotes the number of basis functions used in spatial discretization. Query complexities are measured by numbers of the queries to the fast-forwarding implementation of $e^{-\I A t}$ and the input model of $B(t)$. The query complexity of continuous qDRIFT is measured with respect to the oracle $e^{-iB(t)/p_B(t)}$  and the error is measured in the diamond norm of quantum channels, instead of the operator norm of unitaries.} 
    \label{tab:main_result_2}
\end{table}

As an application, qHOP can be used to accelerate time-dependent Hamiltonian simulation in the interaction picture. 
Consider the dynamics 
\begin{equation}\label{eqn:ham_sim_unbounded}
    \I \partial_t \ket{\psi(t)} = (A+B(t)) \ket{\psi(t)}. 
\end{equation}
Here $A\in \mathbb{C}^{2^{n_s}\times 2^{n_s}}$ is a time-independent Hamiltonian operator and $B(t) \in \mathbb{C}^{2^{n_s}\times 2^{n_s}}$ is a time-dependent Hamiltonian. 
We assume that $A$ has large spectral norm and can be fast-forwarded.
One example of wide applications is the electronic structure problem in the real space formulation~\cite{SuBerryWiebeEtAl2021,AnFangLin2021}, where $A$ comes from spatial discretization of the Laplacian operator $-\Delta$ and $B(t)$ is the discretized time-dependent potential. 
Since $H$ can be split into two parts and $A$ is fast-forwardable, Trotter splitting~\cite{ChildsSuTranEtAl2020} can be directly used to simulate \cref{eqn:ham_sim_unbounded}. 
However, the number of Trotter steps still depends on $\|A\|$ or at least norms of the commutators involving $A$.

The dynamics can be simulated in the interaction picture with a time-dependent Hamiltonian 
\begin{equation}\label{eqn:ham_interaction}
    H_I(t) = e^{\I A t} B(t) e^{-\I A t}.
\end{equation}
Note that $\norm{H_I(t)}=\norm{B(t)}\ll \norm{A}$, but $H_I(t)$ oscillates rapidly in time. \cref{tab:main_result_2} summarizes the results of using different methods to simulate the generic dynamics \cref{eqn:ham_sim_unbounded} as well as the digital simulation of the Schr\"odinger equation with a smooth time-independent potential. 
Note that the query complexity of all methods in the interaction picture scales only logarithmically with respect to $\norm{A}$. 
For the Schr\"odinger equation, this means that the query complexity is only proportional to $\log(N)$, where $N$ is the number of grid points. This significantly reduces the overhead caused by spatial discretization, which is known to be one major concern for efficient simulation of electronic structure problems \cite{KivlichanWiebeBabbushEtAl2017}.

\cref{tab:main_result_2} suggests that for the interaction picture Hamiltonian simulation, the performance of qHOP is at least comparable to that of the first order truncated Dyson method and the continuous qDRIFT method. 
However, if the commutator $\norm{[A,B]}$ and the norm of the derivative $\|B'(t)\|$ are small, then qHOP can achieve second order convergence rate.
In fact, more detailed analysis shows that the condition for second order convergence can be weakened to be $\max_{|s-t|\leq h}\norm{ [B(t), e^{\I A (s-t)} B(s) e^{-\I A (s-t)}]} \leq  C_{AB}h$ and the preconstant $C_{AB}$ is small (see \cref{thm:Mag1_interaction,lem:interaction_commutator}). 

For the Schr\"odinger equation, we prove that this is indeed the case  (see \cref{lem:bound_commutator_realspace}), where the constant $C_{AB}$ only depends on the potential $V$ and can thus scale logarithmically with respect to $N$. 
This implies that qHOP exhibits superconvergence property for simulating the Schr\"odinger equation. In numerical analysis, the term ``superconvergence'' refers to the scenario when a method converges faster than the generally expected convergence rate (see e.g. \cite{Wahlbin2006}). 
In physics, this is sometimes attributed to effects of error interference (see e.g. \cite{TranChuSuEtAl2020}).
Compared to the Monte Carlo method, the continuous qDRIFT method and the first order truncated Dyson method, qHOP improves the scaling in both $\epsilon$ and the simulation time $T$.
The superconvergence property is a surprising result, and its proof rests on a careful application of pseudo-differential  calculus (see \emph{e.g.}, \cite{Stein1993,zworski_book} for an introduction). Our analysis
reveals that  $\norm{[V(x), e^{\I s \Delta} V(x) e^{-\I s \Delta}]}$ (here $A=-\Delta$, $B=V$) can be estimated by $\norm{[V, \Op\left( V(x-2ps) \right)]}$, which is determined by the potential $V$ only, and is independent of the discretization. Here $\Op(\cdot)$ stands for the Weyl quantization operator (see \cref{sec:superconvergence}).

\vspace{1em}
\noindent\textbf{Related works:}

In a different context of classical optimal control simulations for quantum systems with multiple coupled degrees of freedom, Ref. \cite{MaMagannHoRabitz2020} recently proposed to simulate the interaction picture Hamiltonian in \cref{eqn:ham_interaction} as $\wt{U}(T,0) = \exp\left(-\I \int_0^T H_I(s) ds\right)$, and the integral is approximated by a numerical quadrature. 
This is the same as qHOP for interaction picture simulation with a single time segment. 
Therefore, it is surprising that even with such a crude approximation, the resulting numerical scheme can still achieve good accuracy for quantum control applications. This can be viewed as supporting evidence of the effectiveness of qHOP for simulating more general quantum dynamics.

The efficiency of the truncated Dyson series also rests on the accurate numerical quadrature for approximating certain integrals. 
As a result, the query complexity depends only logarithmically on $\norm{H'(t)}$. This logarithmic dependence can be removed by considering the permutation expansion based approach \cite{ChenKalevHen2021}, which evaluates the integrals analytically using divided differences, provided that the Hamiltonian can be written in a finite sum of the permutation expansion with its time-dependent components in the form of exponential sums. These methods are particularly appealing when $\norm{H'(t)}$ is large. However, the cost of the truncated Dyson series still depends polynomially on $\norm{H(t)}$, instead of the commutators.

The continuous qDRIFT  method is an intrinsically probabilistic method, i.e. it approximates the unitary evolution in the weak sense, and the accuracy should be measured by the diamond norm in terms of the corresponding quantum channels~\cite{SuBerryWiebeEtAl2021}. 
The cost is also insensitive to $\norm{H'(t)}$ but depends on $\norm{H(t)}$. It is worth noting that an efficient implementation of the continuous qDRIFT method requires \textit{a priori} information of the norm $\norm{H(t)}$ at each time $t$ (relative to the overall $L^1$ norm $\int_0^T \norm{H(t)} dt$). 
In the interaction picture Hamiltonian simulation, $\norm{H_I(t)}$ is a constant with respect to $t$, which facilitates the implementation of the continuous qDRIFT method for interaction picture simulation~\cite{RajputRoggeroWiebe2021}. Another randomized algorithm is the Monte Carlo method proposed in \cite{PoulinQarrySommaEtAl2011}.  It is worth noting that its query complexity has a multiplicative dependence on the number of quadrature points $M$, while ours scales as $\Or(\log M)$ thanks to the efficiency of the LCU procedure.

In order to simulate the dynamics in \cref{eqn:ham_sim_unbounded}, the cost of Trotter methods depend on the norm of the commutator $\norm{[A,B(t)]}$ or that of nested commutators such as $\norm{[A,[A,B(t)]]}$. Such a commutator scaling is already a significant improvement over the polynomial dependence on $\norm{A}$ in the time-independent case~\cite{ChildsSuTranEtAl2020}. 
However, this still leads to the polynomial dependence on $N$ for the digital simulation of the Schr\"odinger equation.  
It is worth noting that
 the $N$-independent error bound can also be achieved for Trotter-type algorithms, if the error is measured in terms of the vector norm rather than the operator norm, and if the initial vector satisfies certain regularity assumptions~\cite{AnFangLin2021}.
On the other hand, the error of qHOP is measured in the operator norm, and therefore is applicable to arbitrary initial vectors.  
It is interesting to observe that in the interaction picture Hamiltonian simulation, if we use the mid-point rule to approximate the integral in qHOP, then we exactly recover the second order Trotter method (see \cref{sec:trotter_connection}).
Our numerical results verify the advantage of qHOP over Trotter methods when applied to oscillatory initial wavepackets.

\section{Preliminaries}\label{sec:prelim}

In this section, we briefly introduce the concept and the properties of the block-encoding that we will use throughout the paper. 
The definition and the results presented here mostly follow the work~\cite{GilyenSuLowEtAl2019}. 

\begin{definition}[Block-encoding]
    Suppose $A$ is a matrix in $\mathbb{C}^{2^{n_s}\times 2^{n_s}}$, $\alpha>0$ such that $\|A\| \leq \alpha$, $\epsilon>0$ and $n_a$ is a non-negative matrix. 
    Then a unitary matrix $U \in \mathbb{C}^{2^{n_s+n_a}\times 2^{n_s+n_a}}$ is an $(\alpha,n_a,\epsilon)$-block-encoding of the matrix $A$, if 
    \begin{equation}
        \left\|A - \alpha \bra{0}^{\otimes n_a} U \ket{0}^{\otimes n_a} \right\| \leq \epsilon. 
    \end{equation}
\end{definition}

Intuitively, the main idea of the block-encoding is to represent the matrix $A$ as the upper-left block of a unitary, \emph{i.e.},
\begin{equation}
    U \approx \left(\begin{array}{cc}
        A/\alpha & * \\
        * & *
    \end{array}\right). 
\end{equation}
Block-encoding is a powerful input model for computing functions of matrices beyond unitaries. 
Although it is not totally clear how to build a block-encoding circuit for an arbitrarily given matrix, there exist efficient approaches to construct block-encodings for a large subset of matrices of practical interest, including unitaries, density operators, POVM operators and sparse-access matrices~\cite{GilyenSuLowEtAl2019}. 
In this work we simply assume that block-encodings of certain matrices are available and take them as our input models. 

Now we discuss the computations of block-encoded matrices. 
In general it is allowed to add, subtract and multiply two block-encoded matrices. 
Smooth functions of a block-encoded Hermitian matrix can also be efficiently implemented using the quantum singular value transformation (QSVT) technique. 
In particular, in this work, we need the multiplication of block-encodings and the implementation of the function $e^{-\I t H}$ for a Hermitian matrix $H$, which can be implemented by first using even and odd polynomials to approximate $\cos(tH)$ and $\I \sin(tH)$ via QSVT, respectively, then combining them to construct a block-encoding of $e^{-\I H t}/2$, and then using robust oblivious amplitude amplification (OAA)~\cite{BerryChildsCleveEtAl2015} to get a block-encoding of $e^{-\I H t}$.  
The corresponding results are summarized in the following two lemmas, of which the proof can be found in~\cite{GilyenSuLowEtAl2019}. 

\begin{lemma}[Multiplication of block-encoded matrices]\label{lem:product_block_encoding}
    For two matrices $A,B\in\mathbb{C}^{2^{n_s}\times 2^{n_s}}$, if $U_A$ is an $(\alpha,n_a,\delta)$-block-encoding of $A$ and $U_B$ is a $(\beta,n_b,\epsilon)$-block-encoding of $B$, then $(I_{n_b}\otimes U_A)(I_{n_a}\otimes U_B)$ is an $(\alpha\beta,n_a+n_b,\alpha\epsilon+\beta\delta)$-block-encoding of $AB$. 
\end{lemma}

\begin{lemma}[Time-independent Hamiltonian simulation via QSVT and OAA]\label{lem:ham_sim_qsvt}
    Let $\epsilon \in (0,1)$, $t = \Omega(\epsilon)$ and let $U$ be an $(\alpha,n_a,0)$-block-encoding of a time-independent Hamiltonian $H$. 
    Then a unitary $V$ can be implemented such that $V$ is a $(1,n_a+2,\epsilon)$-block-encoding of $e^{-\I t H}$, with $\Or\left(\alpha t + \log(1/\epsilon)\right)$ uses of $U$, its inverse or controlled version,  $\Or\left(n_a(\alpha t + \log(1/\epsilon))\right)$ two-qubit gates and $\mathcal{O}(1)$ additional ancilla qubits. 
\end{lemma}

\section{Quantum highly oscillatory protocol}

In this section, we first show how to derive the qHOP for general time-dependent Hamiltonian simulation of \cref{eqn:ham_sim}. 
Our method can be established by truncating the Magnus expansion~\cite{Magnus1954} to the first order. 
A major difference of the qHOP from the classical Magnus methods~\cite{Thalhammer2006} is that qHOP can estimate the integral of fast oscillatory function in a high accuracy with low cost, as being elaborated later in this section. 
Then we show how qHOP can be applied to simulate Hamiltonian which can be written as the sum of a fast-forwardable unbounded part and a bounded part, in which the original Hamiltonian in the Schr\"odinger picture is transformed to the interaction picture and the corresponding Hamiltonian becomes a bounded time-dependent one with fast oscillations. 

\subsection{qHOP for time-dependent Hamiltonian simulation}
\label{sec:qhop_derivation}

\subsubsection{Input model}

In this work we use the same input model as that in~\cite{LowWiebe2019}. 
Assume that we are given the unitary oracle HAM-T which encodes the Hamiltonian evaluated at different discrete time steps. 
More specifically, given the time-dependent Hamiltonian $H(t)$ with $\|H(t)\| \leq \alpha$, two non-negative integers $j, M$, and a time step size $h$, let $\text{HAM-T}_j$ be an $(n_s+n_a+n_m)$-qubit unitary oracle with $n_m = \log_2 M$ such that 
\begin{equation}
    \bra{0}_a\text{HAM-T}_j\ket{0}_a = \frac{1}{\alpha} \sum_{k=0}^{M-1} \ket{k}\bra{k} \otimes H(jh+kh/M). 
\end{equation}
Here $M$ is the number of nodes used in numerical integration, $h$ is the time step size in the time discretization, $j$ represents the current local time step, and $n_s,n_a$ denote the number of the state space qubits and ancilla qubits, respectively. 
The meanings of the parameters will be further clarified later. 

\subsubsection{Derivation of the method}

We now derive qHOP for simulating the dynamics \cref{eqn:ham_sim}. 
The exact evolution operator of \cref{eqn:ham_sim} can be represented as 
\begin{equation}
    U(T,0) = \mathcal{T} e^{-\I \int_0^T H(s) ds}
\end{equation}
where $\mathcal{T}$ is the time-ordering operator. 
In order to discretize and approximate the exact evolution operator, we first divide the entire time interval $[0,T]$ into $L$ equi-length segments. 
Let the time step size $h = T/L$, then 
\begin{equation} \label{eqn:Ut_product_of_short_time}
    U(t,0) = \prod_{j=0}^{L-1} U((j+1)h,jh) = \prod_{j=0}^{L-1} \mathcal{T} e^{-\I \int_{jh}^{(j+1)h} H(s)ds}. 
\end{equation}
On each segment, the time-ordered evolution operator can be approximated by truncating the Magnus expansion. 
Specifically, Magnus expansion~\cite{Magnus1954,IserlesNorsett1999} tells that, for sufficiently small $h$ such that $h\alpha \leq 1$ \footnote{We remark that although the construction of qHOP can be interprted as truncating the Magnus expansion, effectiveness of qHOP does not require the time step size to be very small because in the error analysis we do not use Magnus expansion. This will be demonstrated in \cref{sec:complexity}.}, we have 
\begin{equation}
    U((j+1)h,jh) = e^{\Omega((j+1)h,jh) }. 
\end{equation}
Here 
\begin{equation}
    \Omega(s,t) = \sum_{k=1}^{\infty} \Omega_k(s,t)
\end{equation}
with 
\begin{equation}
    \Omega_1(s,t) = -\I \int_t^s H(\tau) d\tau 
\end{equation}
and for $k \geq 2$ 
\begin{equation}
    \Omega_k(s,t) = -\I \sum_{l=1}^{k-1} \frac{B_l}{l!} \sum_{p_1+\cdots+p_l = k-1, p_1\geq 1,\cdots,p_l \geq 1} \int_t^s \ad_{\Omega_{p_1}(\tau,t)} \cdots \ad_{\Omega_{p_l}(\tau,t)} H(\tau) d\tau
\end{equation}
where $B_l$ are the Bernoulli numbers. 
Approximating $\Omega(s,t)$ by the single first order term $\Omega_1(s,t)$ gives the approximation
\begin{equation} \label{eqn:U_short_no_timeordering}
     U((j+1)h,jh) \approx e^{-\I \int_{jh}^{(j+1)h}H(s)ds}. 
\end{equation}
Notice that this is equivalent to directly ignoring the time-ordering operator. 
The integral can be further approximated using standard first order quadrature~\cite{BurdenNA} with $M$ nodes as 
\begin{equation} \label{eqn:int_quadrature}
    \int_{jh}^{(j+1)h} H(s)ds  \approx  \frac{h}{M}\sum_{k=0}^{M-1}H(jh+(kh/M)). 
\end{equation}
Then the short-time qHOP can be written as 
\begin{equation}\label{eqn:Mag1}
    U_1((j+1)h,jh) = e^{-\I h \frac{1}{M}\sum_{k=0}^{M-1}H(jh+(kh/M))}. 
\end{equation}
Long-time evolution can thereby be approximated by the multiplication of short-time qHOP evolution operator as 
\begin{equation}
    U(T,0) \approx \prod_{j=0}^{L-1} U_1((j+1)h,jh). 
\end{equation}

\begin{figure}
    \centerline{
    \Qcircuit @R=1em @C=1em {
    \text{Ancilla}\quad\quad\quad\quad & \qw & \multigate{2}{\text{HAM-T}_j} & \qw  & \qw \\
    \text{Control}\quad\quad\quad\quad & \gate{\otimes_m\text{HAD}} & \ghost{\text{HAM-T}_j} & \gate{\otimes_m\text{HAD}} &  \qw \\
    \text{State}\quad\quad\quad\quad & \qw  & \ghost{\text{HAM-T}_j} & \qw  & \qw  \\
    }
    }
    \caption{Quantum circuit of implementing a block-eocoding of the Hamiltonian formulated in \cref{eqn:Mag1_LCU}. The short-time qHOP evolution operator can then be implemented according to \cref{lem:ham_sim_qsvt} using the circuit described here as the input block-encoding model. Here $\text{HAD}$ represents the single qubit Hadamard gate. }
    \label{fig:qHOP_circuit}
\end{figure}
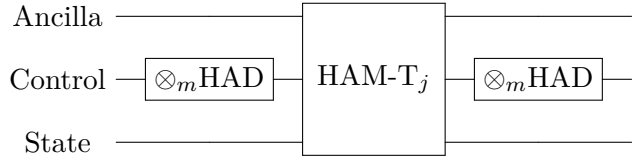

The unitary operator $U_1$ can be simply implemented on a quantum computer by using a particular case of the linear combination of unitary technique~\cite{ChildsWiebe2012} with HAM-T as select oracle and the quantum singular value transform technique for Hamiltonian simulation~\cite{GilyenSuLowEtAl2019}. 
More precisely, by applying $\otimes_m \text{HAD}$ on the $n_m$ qubits where $\text{HAD}$ represents the single qubit Hadamard gate, applying HAM-T and then uncomputing, a block-encoding of the quadrature formula can be constructed such that 
\begin{align}
    & \quad \left(\bra{0}_a \otimes \bra{0}_m \right) \left(I_a\otimes (\otimes_m\text{HAD})\otimes I_s \right) \text{HAM-T}_j \left(I_a\otimes (\otimes_m\text{HAD})\otimes I_s \right) \left(\ket{0}_a \otimes \ket{0}_m \right) \nonumber \\
    & = \frac{1}{M \alpha} \sum_{k=0}^{M-1} H(jh+kh/M) .\label{eqn:Mag1_LCU} 
\end{align}
The quantum circuit of implementing \cref{eqn:Mag1_LCU} is described in \cref{fig:qHOP_circuit}. 
We then implement Hamiltonian simulation of this block-encoding matrix with time $h$ using the result of \cref{lem:ham_sim_qsvt}, then it gives a block-encoding of $U_1((j+1)h,jh)$. 
Finally, the long-time qHOP evolution operator can be block-encoded by the multiplication of the block-encodings $U_1((j+1)h,jh)$ for $j$ from $0$ to $(M-1)$. 

While we leave the rigorous complexity analysis to the next section, we would like to briefly explain why the cost of qHOP is not severely affected by the fast oscillation within $H(t)$. 
This is because the approximation error of truncating the Magnus expansion is independent of $H'(t)$ and, although the number of the quadrature nodes $M$ depends polynomially on $\|H'(t)\|$, the cost of the linear combination of $M$ matrices scales logarithmically in $M$. 
Such an accurate implementation of the numerical integral for fast oscillatory function with poly-logarithmic cost is the key reason why qHOP can significantly reduce the overhead brought by the oscillations, and we are not aware of any classical analog of this feature. 
In particular, generic classical numerical solvers for fast oscillatory differential equation, such as Runge-Kutta method, multistep method and classical Magnus method, can only approximate the integral using a small number of quadrature nodes and thus have polynomial cost in terms of $\|H'(t)\|$. 
We remark that the accurate quantum implementation of the numerical quadrature is also a key component of the truncated Dyson method, of which the cost depends poly-logarithmically on $\|H'(t)\|$.

\subsection{qHOP for unbounded Hamiltonian simulation in the interaction picture}

As an application of the general qHOP, we now discuss how to apply qHOP to simulate Hamiltonian specified in \cref{eqn:ham_sim_unbounded} with large but fast-forwardable $A$ and bounded $B(t)$. 
This idea is to first transfer the dynamics into the interaction picture with the resulting time-dependent Hamiltonian $H_I(t)$, which is bounded but oscillates rapidly in time. 
This regime can be efficiently handled by qHOP.

\subsubsection{Input model}

We assume access to a fast-forwarded Hamiltonian simulation subroutine for the matrix $A$ with a large spectral radius, and the HAM-T oracle for $B(t)$. 
Specifically, we assume that the following two oracles are given: 
\begin{enumerate}
    \item $O_A(s)$ which can fast-forward $e^{\I A s}$ for any $s \in \mathbb{R}$. 
    \item $O_B(j)$, which is the HAM-T oracle for $B(t)$ on the interval $[jh,(j+1)h]$, namely an $(n_s+n_B+n_m)$-qubit unitary oracle with $n_m = \log_2 M$ and $n_B$ denoting the number of ancilla qubits such that\footnote{With some slight abuse of notation, the subscript $a$ is short for ancilla, and the number of the ancilla qubits is $n_B$. } 
    \begin{equation}\label{eqn:HAM-T_B_interaction}
        \bra{0}_a O_B(j) \ket{0}_a = \sum_{k=0}^{M-1} \ket{k}\bra{k} \otimes \frac{B(jh+kh/M)}{\alpha_B}.
    \end{equation}
    Here $\alpha_B$ is the block-encoding factor such that $\max_{t\in[0,T]}\|B(t)\|\leq \alpha_B$. 
\end{enumerate}

\subsubsection{Interaction picture}

To avoid possible polynomial complexity dependence on the large norm $\|A\|$, we need first transform to the interaction picture. 
Let 
\begin{equation}
    \ket{\psi_I(t)} = e^{\I A t}\ket{\psi(t)},
\end{equation}
then 
\begin{equation}\label{eqn:ham_sim_interaction}
    \I \partial_t \ket{\psi_I(t)} = H_I(t) \ket{\psi_I(t)}, 
\end{equation}
where
\begin{equation}\label{eqn:ham_interaction_2}
    H_I(t) = e^{\I A t} B(t) e^{-\I A t}. 
\end{equation}
Notice that \cref{eqn:ham_interaction_2} describes a bounded Hamiltonian. 
However, it becomes time-dependent and its derivative still depends on the norm of the matrix $A$. 
The exact evolution operator of \cref{eqn:ham_sim_interaction} is given as 
\begin{equation}
    U(t,0) = \mathcal{T} e^{-\I \int_0^t H_I(s)ds}.
\end{equation}

\subsubsection{qHOP for interaction picture simulation}

After the dynamics is formulated in the interaction picture, local qHOP evolution operator in \cref{eqn:Mag1} can readily  be  applied, which leads to the operator
\begin{equation} \label{eqn:Mag1_IP_u1_quadrature}
    U_1((j+1)h,jh) = e^{-\I h \frac{1}{M}\sum_{k=0}^{M-1}H_I(jh+kh/M)}. 
\end{equation}
According to the definition of $H_I$, we can further plug the equation $H_I(jh+kh/M) = e^{\I A jh} e^{\I A kh/M} B(jh+kh/M) e^{-\I A kh/M} e^{-\I A jh}$ into the propagator and obtain 
\begin{equation}\label{eqn:Mag1_interaction}
    U_1((j+1)h,jh) = e^{\I A jh}\exp\left(-\I h \frac{1}{M}\left(\sum_{k=0}^{M-1} e^{\I A kh/M} B(jh+kh/M) e^{-\I A kh/M}\right) \right)e^{-\I A jh}. 
\end{equation}
The reason for rewriting the qHOP propagator is that within the summation in \cref{eqn:Mag1_interaction} we only need to implement $e^{-\I A s}$ for $|s| \leq h$. 
Therefore, the evolution time of $A$ is independent of the choice of $j$, and thus the controlled evolution operator regarding $A$ is the same for every step of propagation.

\begin{figure}
    \centerline{
    \Qcircuit @R=1em @C=1em {
    \text{Control}\quad\quad\quad\quad  & \gate{\otimes_m\text{HAD}} & \ctrl{2} & \multigate{2}{O_B(j)} & \ctrl{2} & \gate{\otimes_m\text{HAD}} &  \qw \\
    \text{Ancilla}\quad\quad\quad\quad  & \qw & \qw & \ghost{O_B(j)}& \qw & \qw &  \qw \\
    \text{State}\quad\quad\quad\quad & \gate{O_A(-jh)} & \gate{O_A\left(-\frac{kh}{M}\right)} & \ghost{O_B(j)} & \gate{O_A\left(\frac{kh}{M}\right)} & \gate{O_A(jh)} & \qw  \\
    }
    }
    \caption{ Quantum circuit of implementing the block-encoding of the linear combination of interaction picture Hamiltonians for $H = A+B(t)$. The short-time qHOP evolution operator can then be implemented according to \cref{lem:ham_sim_qsvt} using the circuit here as the input block-encoding of the Hamiltonian. Here $\text{HAD}$ represents the single qubit Hadamard gate.  Here $k$ is a dummy variable used in \cref{eqn:HAM-T_interaction_picture}.}
    \label{fig:qHOP_circuit_interaction_picture}
\end{figure}
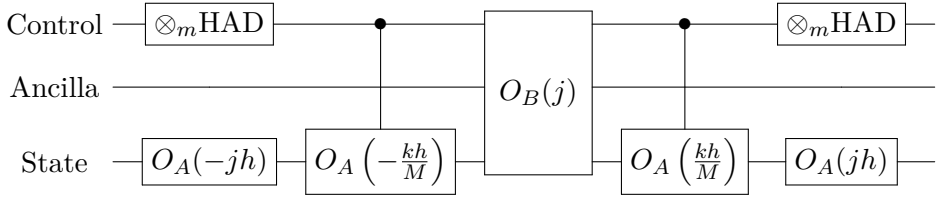

Now we describe how to implement qHOP for interaction picture Hamiltonian simulation. 
Here we only focus on the procedure of constructing the circuit, and we leave the error and complexity analysis to the next section. 
First, the HAM-T oracle encoding $H_I(t)$ can be constructed following the procedure detailed in~\cite{LowWiebe2019}. 
Specifically, according to the binary encoding of each $0 \leq k < M$, we can use  controlled unitaries of $e^{\I A h/M}, e^{\I A 2h/M}, e^{\I A 4h/M}, \cdots, e^{\I A 2^{\log_2(M)}h/M}$ to implement the controlled-evolution operator 
\begin{equation}
    R_A = \sum_{k=0}^{M-1} \ket{k}\bra{k} \otimes e^{\I A kh/M}. 
\end{equation}
Next, we multiply it on the right with $O_B(j)$ and the adjoint of the previous controlled-evolution operator, then apply $e^{\I Ajh}$ and $e^{-\I Ajh}$ at the beginning and the end, we have the $\text{HAM-T}_j$ oracle as
\begin{align}
    & \quad \bra{0}_a  \text{HAM-T}_j \ket{0}_a \nonumber \\
    & =  \bra{0}_a \left(I_a\otimes I_m \otimes e^{\I A jh}\right)\left(I_a\otimes R_A\right)O_B(j) \left(I_a\otimes R_A^{\dagger}\right)\left(I_a\otimes I_m \otimes e^{-\I A jh}\right) \ket{0}_a \nonumber \\ 
    &= \left(I_m \otimes e^{\I A jh}\right)\left(\sum_{k=0}^{M-1} \ket{k}\bra{k} \otimes \frac{e^{\I A kh/M} B(jh+kh/M) e^{-\I A kh/M}}{\alpha_B}\right) \left(I_m \otimes e^{-\I A jh}\right) \\
    &= \sum_{k=0}^{M-1} \ket{k}\bra{k} \otimes \frac{H_I(jh+kh/M)}{\alpha_B} \label{eqn:HAM-T_interaction_picture}. 
\end{align}
Then, the same as the general scenario, the linear combination of the interaction Hamiltonian can be block-encoded as 
\begin{align}
    & \quad \left(\bra{0}_a \otimes \bra{0}_m \right) \left(I_a\otimes (\otimes_m\text{HAD})\otimes I_s \right) \text{HAM-T}_j \left(I_a\otimes (\otimes_m\text{HAD})\otimes I_s \right) \left(\ket{0}_a \otimes \ket{0}_m \right) \nonumber \\
    & = \frac{1}{M \alpha_B } \sum_{k=0}^{M-1} H_I(jh+kh/M). \label{eqn:LCU_interaction}
\end{align}
The entire quantum circuit for constructing a block-encoding for such a linear combination of interaction picture Hamiltonian is summarized in \cref{fig:qHOP_circuit_interaction_picture}. 
Finally, according to \cref{lem:ham_sim_qsvt}, the short time evolution operator $U((j+1)h,jh)$ can be implemented using this circuit as the input block-encoded Hamiltonian. 
The long-time qHOP evolution operator can then be block-encoded by the multiplication of the short-time block-encoding qHOP operators. 

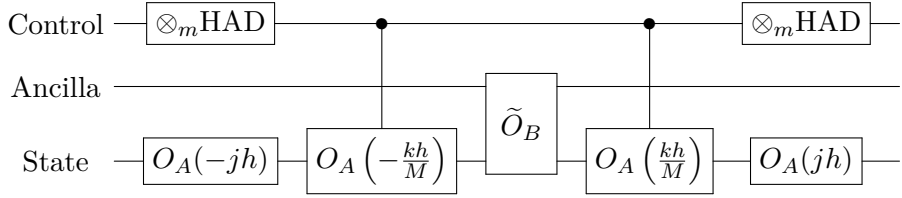
\begin{figure}
    \centerline{
    \Qcircuit @R=1em @C=1em {
    \text{Control}\quad\quad\quad\quad  & \gate{\otimes_m\text{HAD}} & \ctrl{2} & \qw & \ctrl{2} & \gate{\otimes_m\text{HAD}} &  \qw \\
    \text{Ancilla}\quad\quad\quad\quad  & \qw & \qw & \multigate{1}{\widetilde{O}_B}& \qw & \qw &  \qw \\
    \text{State}\quad\quad\quad\quad & \gate{O_A(-jh)} & \gate{O_A\left(-\frac{kh}{M}\right)} & \ghost{\widetilde{O}_B} & \gate{O_A\left(\frac{kh}{M}\right)} & \gate{O_A(jh)} & \qw  \\
    }
    }
    \caption{ Quantum circuit of implementing the block-encoding of the linear combination of interaction picture Hamiltonians for $H = A+B$, which is a special case of \cref{fig:qHOP_circuit_interaction_picture} with a time-independent $B$. Here $\widetilde{O}_B$ represents a block-encoding of the matrix $B$. Here $\text{HAD}$ represents the single qubit Hadamard gate.  Here $k$ is a dummy variable used in \cref{eqn:HAM-T_interaction_picture}.}
    \label{fig:qHOP_circuit_interaction_picture_time_independent}
\end{figure}

As a special case, when the matrix $B(t) \equiv B$ is time-independent, the input model can be simplified. 
Specifically,  the right hand side of \cref{eqn:HAM-T_B_interaction}  becomes $I_m \otimes (B/\alpha_B)$. The $O_B(j)$ oracle becomes a standard block-encoding of the matrix $B$, and the  $n_m$  ancilla qubits are no longer needed to implement $O_B(j)$. 
Therefore, in this case, we can change the input model for $B$ to an $(\alpha_B,n_B,0)$-block-encoding oracle, denoted by $\widetilde{O}_B$. 
The qHOP evolution operator can then be constructed in the same way as the time-dependent case only with replacing $O_B(j)$ there by $\widetilde{O}_B$. 
For clarity, we give the circuit for implementing a linear combination of $H_I$ with time-independent $B$ in \cref{fig:qHOP_circuit_interaction_picture_time_independent}.

\subsubsection{Connection to Trotter formulae}\label{sec:trotter_connection}

It is worth noting that in the context of  interaction picture simulation, qHOP naturally generalizes the Trotter formulae. In particular, when a quantum protocol for the quadrature was not applied, the application of the mid-point quadrature rule provides the second-order Trotter formula, while the first-order Trotter formula correspond to the end-point quadrature rule.
For simplicity here we restrict our discussion with a time-independent matrix $B$. 

As mentioned in the construction of qHOP \cref{eqn:Ut_product_of_short_time} and \cref{eqn:U_short_no_timeordering}, the exact dynamics is first approximated by the dynamics without time-ordering
\begin{align*} \label{eqn:ip_no_time_order}
    & e^{-\I A L h} \mathcal{T} e^{-\I \int_{(L-1)h}^{L h} H_I(s) \, ds} \cdots  \mathcal{T}e^{-\I \int_{0}^{ h} H_I(s) \, ds}
    \\
    \approx & e^{-\I A L h} e^{-\I \int_{(L-1)h}^{L h} H_I(s) \, ds} \cdots  e^{-\I \int_{0}^{ h} H_I(s) \, ds},
\end{align*}
where $Lh = T$ and then followed by a \textit{quantum} numerical quadrature. We now apply the midpoint quadrature rule instead~\cite{BurdenNA}
\[
\int_a^b f(x) \,dx \approx f\left( (a+b)/2 \right) (b-a),
\]
and obtain
\begin{align*}
    & e^{-\I A Lh} e^{-\I H_I\left(Lh-h/2\right) h}  \cdots  e^{-\I  H_I(h/2) h} 
    \\
    = & e^{-\I A Lh} e^{-\I e^{\I A (Lh - h/2)} B e^{-\I A (Lh - h/2 )} h}  \cdots  e^{-\I e^{\I A  h/2} B e^{-\I A  h/2} h }
    \\
    = & e^{-\I A Lh} e^{\I A (Lh - h/2)}  e^{-\I  B h}   e^{-\I A (Lh - h/2 )}  \cdots e^{\I A  h/2}  e^{-\I  B h} e^{-\I A  h/2}
    \\
    = & e^{-\I A h/2} e^{-\I  B h}   e^{-\I A h} e^{-\I B h}  \cdots e^{-\I A  h}  e^{-\I  B h} e^{-\I A  h/2},
\end{align*}
which recovers the second-order Trotter formula. Similarly, the first-order Trotter formula can be derived via the end-point quadrature rule
\[
\int_a^b f(x) \,dx \approx f\left( a\right) (b-a),
\]
and we do not detail here.

\section{Complexity analysis}\label{sec:complexity}

In this section we study the complexity of qHOP to obtain an $\epsilon$-approximation of the exact evolution operator up to time $T$. 
We first analyze the complexity of qHOP for general time-dependent Hamiltonian simulation \cref{eqn:ham_sim}, then study the scenario of Hamiltonian simulation in the interaction picture \cref{eqn:ham_sim_interaction}. 

\subsection{General complexity} 

The proof of the theorem relies on the error bound of qHOP, which can be further decomposed into two parts: the time discretization error and the error in constructing the block-encodings. 
We will first establish the time discretization error in a single time step, then use this error bound to analyze the complexity of constructing short-time qHOP evolution operator, and finally study the long-time scenario. 

\subsubsection{Time discretization errors}

The error of the time discretization can be established by combining standard error bounds of the classical Magnus method and numerical quadrature. 

\begin{lemma}[Time discretization errors of qHOP]\label{lem:Mag1_discretization_error}
    Let $U((j+1)h,jh)$ denote the exact evolution operator $\mathcal{T}e^{-\I \int_{jh}^{(j+1)h}H(s)ds}$, and $U_1((j+1)h,jh)$ denote the qHOP operator defined in \cref{eqn:Mag1}. Then we have 
    \begin{align}
        & \quad \left\|U((j+1)h,jh) - U_1((j+1)h,jh)\right\| \nonumber \\
        &\leq \frac{T^2}{2L^2} \left(\frac{1}{2}\max_{s,\tau \in [jh,(j+1)h]} \left\|[H(\tau),H(s)]\right\| + \frac{1}{M} \max_{s\in[jh,(j+1)h]} \|H'(s)\|\right). 
    \end{align}
\end{lemma}
\begin{proof}
    We first study the approximation error brought by truncating the Magnus expansion.
    Let 
    \begin{equation}
        \widetilde{U}_1(t,jh) = e^{-\I\int_{jh}^t H(s) ds}. 
    \end{equation}
    For any $t \in [jh,(j+1)h]$, by differentiating $\widetilde{U}_1(t,jh)$ with respect to $t$, we have 
    \begin{equation}
        \I \partial_t \widetilde{U}_1(t,jh) = \int_0^1 e^{-\I \beta \int_{jh}^t H(s)ds }  H(t) e^{\I \beta \int_{jh}^t H(s)ds } d \beta \widetilde{U}_1(t,jh). 
    \end{equation}
    By the variation of parameters formula~\cite{HairerNorsettWanner1987,Knapp2005}, we have 
    \begin{align}
        \widetilde{U}_1(t,jh) - U(t,jh) = \int_{jh}^t U(t,\tau) \left(\int_0^1 e^{-\I \beta \int_{jh}^{\tau} H(s)ds }  H(\tau) e^{\I \beta \int_{jh}^{\tau} H(s)ds } d \beta - H(\tau)\right) d\tau. 
    \end{align}
    Using fundamental theorem of calculus in terms of $\beta$ that 
    \begin{align}
        & \quad \left\|e^{-\I \beta \int_{jh}^{\tau} H(s)ds }  H(\tau) e^{\I \beta \int_{jh}^{\tau} H(s)ds } - H(\tau)\right\| \nonumber \\
        & = \left\|\I \int_0^{\beta} e^{-\I \gamma \int_{jh}^{\tau} H(s)ds }  \left[H(\tau), \int_{jh}^{\tau} H(s)ds\right] e^{\I \gamma \int_{jh}^{\tau} H(s)ds } d\gamma \right\| \nonumber \\
        & \leq \beta \int_{jh}^{\tau} \left\|[H(\tau),H(s)]\right\| ds, 
    \end{align}
    we obtain 
    \begin{align}
        \left\|\widetilde{U}_1((j+1)h,jh)-U((j+1)h,jh)\right\| & \leq \int_{jh}^{(j+1)h}\int_0^1 \beta \int_{jh}^{\tau} \left\|[H(\tau),H(s)]\right\| ds d\beta d\tau \nonumber \\
        & = \frac{1}{2} \int_{jh}^{(j+1)h}\int_{jh}^{\tau} \left\|[H(\tau),H(s)]\right\| ds d\tau \nonumber \\
        & \leq \frac{h^2}{4} \max_{s,\tau \in [jh,(j+1)h]} \left\|[H(\tau),H(s)]\right\|.  \label{eqn:error_mag_expansion}
    \end{align}
    
    The quadrature error can be bounded by standard result~\cite{BurdenNA} such that 
    \begin{equation}
        \left\|\int_{jh}^{(j+1)h} H(s) ds - \frac{h}{M}\sum_{k=0}^{M-1}H(jh+kh/M)\right\| \leq \frac{h^2}{2M} \max_{s\in[jh,(j+1)h]} \|H'(s)\|. 
    \end{equation}
    Therefore, by the inequality $\|e^{-\I H_1}-e^{-\I H_2}\| \leq \left\|H_1-H_2\right\|$ for two Hermitian matrices $H_1$ and $H_2$~\cite{ChakrabortyGilyenJeffery2018}, we have 
    \begin{equation}\label{eqn:error_mag_integral}
        \left\|\widetilde{U}_1((j+1)h,jh) - U_1((j+1)h,jh)\right\| \leq \frac{h^2}{2M} \max_{s\in[jh,(j+1)h]} \|H'(s)\|, 
    \end{equation}
    and thus 
    \begin{align}
        & \quad \left\|U((j+1)h,jh) - U_1((j+1)h,jh)\right\| \nonumber \\
        & \leq \left\|U((j+1)h,jh) - \widetilde{U}_1((j+1)h,jh)\right\| + \left\|\widetilde{U}_1((j+1)h,jh) - U_1((j+1)h,jh)\right\| \nonumber \\
        & \leq \frac{h^2}{4} \max_{s,\tau \in [jh,(j+1)h]} \left\|[H(\tau),H(s)]\right\| + \frac{h^2}{2M} \max_{s\in[jh,(j+1)h]} \|H'(s)\| \nonumber \\
        & = \frac{T^2}{2L^2} \left(\frac{1}{2}\max_{s,\tau \in [jh,(j+1)h]} \left\|[H(\tau),H(s)]\right\| + \frac{1}{M} \max_{s\in[jh,(j+1)h]} \|H'(s)\|\right). 
    \end{align}
\end{proof}

Note that $M$ can be chosen to be sufficiently large such that the second part in the error bound becomes negligible, and the additional cost is $\Or(\log M)$. 
The complexity of qHOP is then significantly influenced by the commutator $\max_{s,\tau \in [jh,(j+1)h]} \left\|[H(\tau),H(s)]\right\|$. 
This can be trivially bounded by $2\alpha^2$, which becomes the error bound of the first order truncated Dyson method.  
Furthermore, in many cases of practical interests, the scaling of the commutator can be much better in terms of $\|H\|,\|H'\|,h$, or even a combination of several parameters, which demonstrates the advantage of qHOP over first order truncated Dyson method. 
For technical simplicity, we will first assume an upper bound that $\max_{s,\tau \in [jh,(j+1)h]} \left\|[H(\tau),H(s)]\right\| \leq C_H h^{\theta}$ in the next subsection to establish the complexity estimate in the general case, and specify the parameters $C_H$ and $\theta$ in different scenarios after the generic complexity estimate.

\subsubsection{Short-time and Long-time complexity}

Now we are ready to estimate the complexity scaling of qHOP. 
We first estimate the cost of constructing a block-encoding of short-time evolution operator, then analyze the global cost for long-time simulation.

\begin{lemma}[Short-time complexity of qHOP]\label{lem:Mag1_local_cost}
    Assume that $\max_{s,\tau \in[jh,(j+1)h]} \left\|[H(\tau),H(s)]\right\| \leq C_H h^\theta$ for a non-negative real number $\theta$ and a constant $C_H$ which might depend on $H$. 
    Then for any $0 < \delta < h$, qHOP gives a $(1,n',\delta')$-block-encoding of $U((j+1)h,jh)$ with 
    \begin{align}
        &n' = n_a + \log_2 M + 2, \quad M = 2 \frac{\max_{s\in [0,T]}\|H'(s)\| }{ C_H h^{\theta}}, \\
        &\delta' = \frac{\delta}{2}+ \frac{1}{2} C_H h^{2+\theta}, 
    \end{align}
    and the following cost: 
    \begin{enumerate}
        \item $\mathcal{O}(\alpha h + \log(1/\delta))$ uses of $\text{HAM-T}_j$, its inverse or controlled version, 
        \item $\mathcal{O}((n_a+\log M)(\alpha h + \log(1/\delta)))$ one- or two-qubit gates, 
        \item $\mathcal{O}(1)$ additional ancilla qubits.
    \end{enumerate}
\end{lemma}
\begin{proof}
    We start with \cref{eqn:Mag1_LCU}, which is a $(\alpha,n_a+n_m,0)$-block-encoding of $M^{-1}\sum_{k=0}^{M-1}H(jh+kh/M)$ with $1$ query to $\text{HAM-T}_j$ and $\mathcal{O}(n_m)$ one-qubit gates. 
    According to \cref{lem:ham_sim_qsvt}, a $(1,n_a+n_m+2,\delta/2)$-block-encoding of $U_1((j+1)h,jh)$ can then be implemented by QSVT, with $\mathcal{O}(\alpha h + \log(1/\delta))$ uses of $\text{HAM-T}_j$, its inverse or controlled version, $\mathcal{O}((n_a+n_m)(\alpha h + \log(1/\delta)))$ one- or two-qubit gates, and $\mathcal{O}(1)$ ancilla qubits. 
    
    Now we would like to choose $M$ such that the error of the numerical quadrature becomes less dominant. 
    According to \cref{lem:Mag1_discretization_error}, this can be done by choosing 
    $$M = 2 \frac{\max_{s\in [0,T]}\|H'(s)\| }{ C_H h^{\theta}}. $$ 
    Then the previous circuit becomes a block-encoding of $U((j+1)h,jh)$ with the desired error. 
\end{proof}

Now we are ready to establish the total cost for long-time simulation.

\begin{theorem}[Long-time complexity of qHOP]\label{thm:Mag1_general}
    Let the Hamiltonian $H(s)$ satisfies $\|H(s)\| \leq \alpha$ for all $0\leq s \leq T$, and assume that $\max_{s,\tau \in[jh,(j+1)h]} \left\|[H(\tau),H(s)]\right\| \leq C_H h^\theta$ for a non-negative real number $\theta$ and a constant $C_H$ which might depend on $H$. 
    Then for any $0 < \epsilon < 1, T > \epsilon$, qHOP can implement an operation $W$ such that $\|W - U(T,0)\| \leq \epsilon$ with failure probability $\mathcal{O}(\epsilon)$ and the following cost: 
    \begin{enumerate}
        \item $\mathcal{O}\left(\alpha T + \frac{C_H^{1/(1+\theta)} T^{1+1/(1+\theta)} }{\epsilon^{1/(1+\theta)}}\log\left(\frac{C_H T}{\epsilon}\right)\right)$ uses of $\text{HAM-T}_j$, its inverse or controlled version, 
        \item $\mathcal{O}\left(\left(n_a + \log\left(\frac{\max_{s\in[0,T]} \|H'(s)\| T}{C_H \epsilon }\right)\right)\left(\alpha T + \frac{C_H^{1/(1+\theta)} T^{1+1/(1+\theta)} }{\epsilon^{1/(1+\theta)}}\log\left(\frac{C_H T}{\epsilon}\right)\right)\right)$ one- or two-qubit gates, 
        \item $\mathcal{O}(1)$ additional ancilla qubits. 
    \end{enumerate}
\end{theorem}
\begin{proof}
    The idea of the proof mostly follows the proof of~\cite[Corollary 4]{LowWiebe2019}. 
    Let $\delta' = \delta/2+ C_Hh^{2+\theta}/2$, $n_{b} = n_a + \log_2 M + 2$ and $M = 2\max_{s\in[0,T]}\|H'(s)\|/(C_H h^{2+\theta})$. 
    Let $V_j$ denote the circuit constructed in \cref{fig:qHOP_circuit}, and $W_j = \bra{0}_{b} V_j \ket{0}_{b}$, then \cref{lem:Mag1_local_cost} tells that 
    \begin{equation}
        \left\|W_j - U((j+1)h,jh)\right\| \leq \delta'. 
    \end{equation}
    Notice that $\|W_j\| \leq 1$ and $\|U((j+1)h,jh)\| = 1$, we have 
    \begin{equation}
        \left\|\prod_{j=0}^{L-1} W_j - U(T,0)\right\| \leq L \delta'. 
    \end{equation}
    Each $W_j$ is obtained by applying $V_j$ on $\ket{0}_{b}$ and then projecting back onto $\bra{0}_{b}$. 
    Since $W_j$ is $\delta'$-close to a unitary operator, the norm of $W_j$ is at least $(1-\delta')$. 
    Therefore the failure probability at each single step is bounded by $1-(1-\delta')^2 \leq 2\delta'$. 
    This leads the global failure probability to be bounded by $1-(1-2\delta')^L = \mathcal{O}(L\delta')$.  
    Therefore, in order to bound the error and the failure probability by $\mathcal{O}(\epsilon)$, we can choose $L\delta' \leq \epsilon$ and thus 
    \begin{equation}
        \frac{L\delta}{2}+\frac{C_H T^{2+\theta}}{2L^{1+\theta}} \leq \epsilon. 
    \end{equation}
    By bounding each term on the left hand side by $\epsilon/2$, it suffices to choose 
    \begin{equation}
        L = \frac{C_H^{1/(1+\theta)} T^{1+1/(1+\theta)} }{\epsilon^{1/(1+\theta)}}, \quad \delta = \frac{\epsilon^{1+1/(1+\theta)}}{C_H^{1/(1+\theta)} T^{1+1/(1+\theta)} } 
    \end{equation}
    and correspondingly 
    \begin{equation}
        M = \frac{2\max_{s\in[0,T]}\|H'(s)\|L^{\theta}}{C_HT^{\theta}} = \frac{2\max_{s\in[0,T]}\|H'(s)\| T^{\theta/(1+\theta)}}{C_H^{1/(1+\theta)}\epsilon^{\theta/(1+\theta)}}
    \end{equation}
    The proof is completed by multiplying the cost in \cref{lem:Mag1_local_cost} by $L$ and then plugging in the choices of $L$ and $\delta$. 
\end{proof}

\subsubsection{Scaling of the commutator}

Now we establish the bound for the commutator, \emph{i.e.} estimation of the parameters $C_{H}$ and $\theta$. 
In particular, there are two possible sets of choices of $(C_H,\theta)$, corresponding to two scenarios of the fast oscillations we have discussed in the introduction.  
We then can obtain two corresponding bounds of the asymptotic complexity. 
One of the bounds is independent of $\|H'(s)\|$ and scales linearly with respect to $1/\epsilon$. 
The other one can give a second order convergence, leading to a quadratic speedup in terms of $\epsilon$.
This is at the expense of introducing a polynomial dependence on the commutator $\|[H'(s),H(\tau)]\|$. 
The final complexity estimate can be viewed as the minimum of these two bounds, which indicates that qHOP in the worst case is still at least comparable to the first order truncated Dyson method, and can automatically achieve better complexity without knowing the source of the oscillations in the Hamiltonian. 

We start with the estimate of the commutator, which can be established in the following lemma. 

\begin{lemma}[Bounds for commutators]\label{lem:general_commutator}
    For any $h > 0$, $0\leq j < M$, we have 
        \begin{equation}
            \max_{s,\tau\in[jh,(j+1)h]} \left\|[H(\tau),H(s)]\right\| \leq \min\left\{\max_{|s-u|\leq h} \left\|[H(u),H(s)]\right\|, \max_{|s-u|\leq h} \left\|[H'(u),H(s)]\right\| h\right\}. 
        \end{equation}
\end{lemma}
\begin{proof}
    It suffices to show that the commutator is bounded by both terms in the bracket on the right hand side. 
    The first bound follows trivially. 
    To prove the second bound, we use the fundamental theorem of calculus to obtain 
    \begin{equation}
        H(\tau) = H(s) + \int_{s}^{\tau} H'(u)du
    \end{equation}
    and thus, 
    \begin{equation}
        \max_{s,\tau\in[jh,(j+1)h]}\left\|[H(\tau),H(s)]\right\| = \max_{s,\tau\in[jh,(j+1)h]}\left\|\left[\int_{s}^{\tau} H'(u)du, H(s) \right]\right\| \leq \max_{|s-u|\leq h} \left\|[H'(u),H(s)]\right\| h. 
    \end{equation}
\end{proof}

\cref{lem:general_commutator} implies two possible sets of parameters $(C_{H}, \theta)$: $C_{H} = \max_{s,u\in [0,T]} \left\|[H(u),H(s)]\right\|$, $\theta = 0$, or $C_{H} = \max_{s,u\in[0,T]} \left\|H'(u),H(s)\right\|, \theta = 1$. 
The following result can then be proved by plugging these two possible choices back to \cref{thm:Mag1_general}. 

\begin{cor}[Long-time complexity of qHOP]\label{cor:Mag1_general}
    Let $\|H(s)\| \leq \alpha$, $\max_{s,u\in[0,T]}\left\|[H(u),H(s)]\right\| \leq \widetilde{\alpha}^2$, and $\max_{s,u\in[0,T]}\left\|[H'(u),H(s)]\right\| \leq \widetilde{\beta}$. 
    Then for any $0 < \epsilon < 1, T > \epsilon$, qHOP can implement an operation $W$ such that $\|W - U(T,0)\| \leq \epsilon$ with failure probability $\mathcal{O}(\epsilon)$ and the following cost: 
    \begin{enumerate}
        \item $\mathcal{O}\left( \alpha T + \min\left\{  \frac{\widetilde{\alpha}^2 T^2 }{\epsilon}\log\left(\frac{\widetilde{\alpha} T}{\epsilon}\right),   \frac{\widetilde{\beta}^{1/2} T^{3/2} }{\epsilon^{1/2}}\log\left(\frac{\widetilde{\beta} T}{\epsilon}\right) \right\}  \right)$
        uses of $\text{HAM-T}_j$, 
        \item $\mathcal{O}\left(\left(n_a+\log\left(\frac{\max_{s \in [0, T]}\|H'(s)\| T}{ \min\left\{\widetilde{\alpha}^2,\widetilde{\beta}\right\}\epsilon }\right)\right)\left(\alpha T + \min\left\{  \frac{\widetilde{\alpha}^2 T^2 }{\epsilon}\log\left(\frac{\widetilde{\alpha} T}{\epsilon}\right),   \frac{\widetilde{\beta}^{1/2} T^{3/2} }{\epsilon^{1/2}}\log\left(\frac{\widetilde{\beta} T}{\epsilon}\right) \right\}\right)\right)$ one- or two-qubit gates, 
        \item $\mathcal{O}(1)$ additional ancilla qubits. 
    \end{enumerate}
\end{cor}

\subsubsection{$L^1$ norm scaling}

In this section, we discuss the $L^1$-norm scaling of qHOP for the long time evolution. 
Note that taking the maximum over the whole time interval in our commutator scaling $\max_{s,t\in[0,T]}\|[H(s),H(t)]\|$ can lead to overly pessimistic results. 
In Eq.~\eqref{eqn:error_mag_expansion}, the short time error indeed follows a local $L^1$ scaling. 
In fact, following the same strategy as continuous qDRIFT~\cite{BerryChildsSuEtAl2020}, we can show that the commutator error bound leads to a global $L^1$-norm scaling of the error.

The idea is to vary time step sizes in the propagation according to the average performance of the Hamiltonian. 
To be specific, let $0 = t_0 < t_1 < \cdots < t_j < \cdots < t_L = T$ where $L$ is the number of time steps and $t_1,\cdots,t_{L-1}$ are chosen such that 
\begin{equation}\label{eqn:L1_time_steps}
    \int_0^{t_1} \norm{H(s)} ds = \cdots  = \int_{t_j}^{t_{j+1}} \norm{H(s)} ds = \frac{1}{L} \int_{0}^{T} \norm{H(s)} ds, \quad 0 \leq j \leq L-1. 
\end{equation}
We approximate the exact evolution operator $U(T,0)$ by $\prod_{j=0}^{L-1} U_1(t_{j+1}, t_j)$ where 
\begin{equation}\label{eqn:Mag1_L1}
    U_1(t_{j+1},t_j) = e^{-\I (t_{j+1}-t_j) \frac{1}{M}\sum_{k=0}^{M-1}H(t_j+k(t_{j+1}-t_j)/M)}. 
\end{equation}
Since  the small time steps in the numerical quadrature are different among different time intervals, we require an adaptive version of the HAM-T oracle, which is an $(n_s+n_a+n_m)$-qubit unitary oracle with $n_m = \log_2 M$ such that 
\begin{equation}
    \bra{0}_a\text{HAM-T}_j\ket{0}_a = \frac{1}{\alpha} \sum_{k=0}^{M-1} \ket{k}\bra{k} \otimes H(t_j+k(t_{j+1}-t_j)/M). 
\end{equation}
Notice that this adaptive version of HAM-T can also be efficiently constructed in the interaction picture following a similar circuit in \cref{fig:qHOP_circuit_interaction_picture_time_independent}.

\begin{lemma}[Time discretization errors of qHOP with $L^1$-norm scaling]\label{lem:Mag1_discretization_error_L1}
    Let $0 = t_0 < t_1 < \cdots < t_j < \cdots < t_L = T$ be chosen such that \cref{eqn:L1_time_steps} holds, $U(t_{j+1},t_j)$ denote the exact evolution operator $\mathcal{T}e^{-\I \int_{t_j}^{t_{j+1}}H(s)ds}$, and $U_1(t_{j+1},t_j)$ denotes the qHOP operator defined in \cref{eqn:Mag1_L1}. Then we have 
    \begin{align}
        & \quad \left\|U((j+1)h,jh) - U_1((j+1)h,jh)\right\| \nonumber \\
        &\leq \frac{1}{L^2} \left(\int_{0}^{T} \norm{H(s)} ds  \right)^2 + \frac{(t_{j+1}-t_j)^2}{2M} \max_{s\in[t_j,t_{j+1}]} \|H'(s)\|. 
    \end{align}
\end{lemma}
\begin{proof}
     This lemma can be proved by a slight modification of the proof of \cref{lem:Mag1_discretization_error}. 
     Following the same notations as those in \cref{lem:Mag1_discretization_error}, according to the second line of \cref{eqn:error_mag_expansion}, we have 
     \begin{align}
        \left\|\widetilde{U}_1(t_{j+1},t_j)-U(t_{j+1},t_j)\right\|
        & \leq \frac{1}{2} \int_{t_j}^{t_{j+1}}\int_{t_j}^{\tau} \left\|[H(\tau),H(s)]\right\| ds d\tau \nonumber \\
        & \leq \int_{t_j}^{t_{j+1}}\int_{t_j}^{t_{j+1}} \left\|H(\tau)\right\|\left\|H(s)\right\| ds d\tau \nonumber \\
        & = \left(\int_{t_j}^{t_{j+1}} \norm{H(s)} ds  \right)^2 \nonumber \\
        & = \frac{1}{L^2} \left(\int_{0}^{T} \norm{H(s)} ds  \right)^2. 
    \end{align}
    Combining this estimate with \cref{eqn:error_mag_integral}, we obtain 
    \begin{align}
        & \quad \left\|U(t_{j+1},t_j) - U_1(t_{j+1},t_j)\right\| \nonumber \\
        & \leq \left\|U(t_{j+1},t_j) - \widetilde{U}_1(t_{j+1},t_j)\right\| + \left\|\widetilde{U}_1(t_{j+1},t_j) - U_1(t_{j+1},t_j)\right\| \nonumber \\
        & \leq \frac{1}{L^2} \left(\int_{0}^{T} \norm{H(s)} ds  \right)^2 + \frac{(t_{j+1}-t_j)^2}{2M} \max_{s\in[t_j,t_{j+1}]} \|H'(s)\|.  
    \end{align}
\end{proof}

Using the same strategy as that in \cref{lem:Mag1_local_cost} and \cref{thm:Mag1_general} with the error bound in $L^1$-scaling as specified in \cref{lem:Mag1_discretization_error_L1}, we can obtain another version of the complexity estimates with $L^1$-norm scaling for short-time and long-time propagation.

\begin{lemma}[Short-time complexity of qHOP with $L^1$-norm scaling]\label{lem:Mag1_local_cost_L1}
    Let $\|H(s)\| \leq \alpha$ and $T^{-1}\int_0^T\|H(s)\|ds \leq \overline{\alpha}$. 
    Then for any $0 < \delta < t_{j+1}-t_j$, qHOP gives a $(1,n',\delta')$-block-encoding of $U(t_{j+1},t_j)$ with 
    \begin{align}
        &n' = n_a + \log_2 M + 2, \quad M = \frac{L^2 \max_{s\in[0,T]}\|H'(s)\|}{2\overline{\alpha}^2}, \\
        &\delta' = \frac{\delta}{2}+ \frac{2\overline{\alpha}^2 T^2}{L^2}, 
    \end{align}
    and the following cost: 
    \begin{enumerate}
        \item $\mathcal{O}(\alpha (t_{j+1}-t_j) + \log(1/\delta))$ uses of $\text{HAM-T}_j$, its inverse or controlled version, 
        \item $\mathcal{O}((n_a+\log M)(\alpha (t_{j+1}-t_j) + \log(1/\delta)))$ one- or two-qubit gates, 
        \item $\mathcal{O}(1)$ additional ancilla qubits.
    \end{enumerate}
\end{lemma}

\begin{theorem}[Long-time complexity of qHOP with $L^1$-scaling]\label{thm:Mag1_general_L1}
    Let $\|H(s)\| \leq \alpha$ for any $0\leq s\leq T$, and $T^{-1}\int_0^T\|H(s)\|ds \leq \overline{\alpha}$. 
    Then for any $0 < \epsilon < 1, T > \epsilon$, qHOP can implement an operation $W$ such that $\|W - U(T,0)\| \leq \epsilon$ with failure probability $\mathcal{O}(\epsilon)$ and the following cost: 
    \begin{enumerate}
        \item $\mathcal{O}\left(\alpha T + \frac{\overline{\alpha}^2 T^2}{\epsilon} \log\left(\frac{\overline{\alpha}T}{\epsilon}\right)\right)$ uses of $\text{HAM-T}_j$, its inverse or controlled version, 
        \item $\mathcal{O}\left(\left(n_a+\log \left( \frac{\overline{\alpha} T \max_{s\in[0,T]}\|H'(s)\|}{\epsilon}\right)\right)\left(\alpha T + \frac{\overline{\alpha}^2 T^2}{\epsilon} \log\left(\frac{\overline{\alpha} T}{\epsilon}\right)\right)\right)$ one- or two-qubit gates, 
        \item $\mathcal{O}(1)$ additional ancilla qubits. 
    \end{enumerate}
\end{theorem}
\begin{proof}
    Following exactly the proof of \cref{thm:Mag1_general}, we need to bound the error and the failure probability by $\mathcal{O}(\epsilon)$ by choosing $L\delta' \leq \epsilon$ and thus 
    \begin{equation}
        \frac{L\delta}{2}+\frac{2\overline{\alpha}^2 T^2}{L} \leq \epsilon. 
    \end{equation}
    By letting each term on the left hand side equal $\epsilon/2$, it suffices to choose 
    \begin{equation}
        L = \frac{4\overline{\alpha}^2 T^2}{\epsilon} , \quad \delta = \frac{\epsilon^2}{4\overline{\alpha}^2 T^2} ,
    \end{equation}
    and correspondingly 
    \begin{equation}
        M = \frac{L^2 \max_{s\in[0,T]}\|H'(s)\|}{2\overline{\alpha}^2} = \frac{8\overline{\alpha}^2 T^4 \max_{s\in[0,T]}\|H'(s)\|}{\epsilon^2}
    \end{equation}
    The proof is completed by taking the summation of the local costs in \cref{lem:Mag1_local_cost_L1} over the entire $[0,T]$ and then plugging in the choices of $L$, $\delta$ and $M$. 
\end{proof}

\subsection{Complexity of the Hamiltonian simulation in the interaction picture}

The complexity estimate of qHOP applied to Hamiltonian simulation in the interaction picture can be obtained by directly applying \cref{thm:Mag1_general} in the generic case. 
The difference is that, under the interaction picture, we can get a more concrete expression of the commutator $\|[H(u),H(s)]\|$ and $\|[H'(u),H(s)]\|$, which leads to improved scaling in various scenarios and examples. 
We will first give the generic complexity estimates for short-time and long-time simulation in the interaction picture.

\subsubsection{Complexity} 
\begin{lemma}[Short-time complexity of qHOP in the interaction picture]\label{lem:Mag1_local_cost_interaction}
    Assume that $\max_{t\in[0,T]}\|B(t)\| \leq \alpha_B$, $\max_{t\in[0,T]}\|B'(t)\| \leq \beta_B$, $\max_{t\in[0,T]}\|[A,B(t)]\| \leq \alpha_{AB}$, and that  $$\max_{|s-t|\leq h }\norm{ [B(t), e^{\I A (s-t)} B(s) e^{-\I A (s-t)}]} \leq  C_{AB}h^{\theta}$$ for a non-negative real number $\theta$ and a constant $C_{AB}$ which might depend on $A$ and $B(t)$. 
    Then for any $0 < \delta < h$, qHOP gives a $(1,n',\delta')$-block-encoding of $U((j+1)h,jh)$ with 
    \begin{align}
        &n' = n_B+2 + \mathcal{O}\left(\log M \right), \quad M = \frac{2(\alpha_{AB}+\beta_B)}{C_{AB} h^{\theta}}, \\
        &\delta' = \delta/2+C_{AB}h^{2+\theta}/2, 
    \end{align}
    and the following cost: 
    \begin{enumerate}
        \item $\mathcal{O}(\left(\alpha_B h + \log(1/\delta)\right) \log M )$ uses of $O_A$, its inverse or controlled version, 
        \item $\mathcal{O}(\alpha_B h + \log(1/\delta))$ uses of $O_B(j)$, its inverse or controlled version, 
        \item $\mathcal{O}((n_B+\log M)(\alpha_B h + \log(1/\delta)))$ one- or two-qubit gates, 
        \item $\mathcal{O}(1)$ additional ancilla qubits.
    \end{enumerate}
\end{lemma}
\begin{proof}
    This is a direct consequence of \cref{lem:Mag1_local_cost}. As discussed in \cref{sec:qhop_derivation}, the circuit for $\text{HAM-T}_j$  can be constructed with $\mathcal{O}(\log M)$ uses of controlled $O_A$ and $\mathcal{O}(1)$ use of $O_B(j)$, 
    \begin{align*}
        \left\|[H_I(t),H_I(s)]\right\| &= \left\|e^{\I A t}B(t)e^{-\I A t} e^{\I A s}B(s)e^{-\I A s} - e^{\I A s}B(s)e^{-\I A s}e^{\I A t}B(t)e^{-\I A t}\right\| \\
        & = \left\|e^{\I A t }\left(B(t) e^{\I A (s-t)}B(s)e^{-\I A (s-t)} - e^{\I A (s-t)}B(s)e^{-\I A (s-t)}B(t)\right) e^{-\I A t } \right\|\\
        & = \left\|e^{\I A t }\left[B(t),e^{\I A (s-t)}B(s)e^{-\I A (s-t)} \right] e^{-\I A t }\right\| \\
        & \leq \left\|\left[B(t),e^{\I A (s-t)}B(s)e^{-\I A (s-t)} \right] \right\|, 
    \end{align*}
    and 
    \begin{equation}
        \left\|H_I'(s)\right\| = \left\|\I e^{\I A s} [A,B(s)] e^{-\I A s} + e^{\I A s} B'(s) e^{-\I A s}\right\| \leq  \left\|[A,B(s)]\right\| + \|B'(s)\|. 
    \end{equation}
\end{proof}

The complexity for long-time simulation directly follows from \cref{thm:Mag1_general} for the same reason. 

\begin{theorem}[Long-time complexity of qHOP in the interaction picture]\label{thm:Mag1_interaction}
    Assume that $\max_{t\in[0,T]}\|B(t)\| \leq \alpha_B$, $\max_{t\in[0,T]}\|B'(t)\| \leq \beta_B$,  $\max_{t\in[0,T]}\|[A,B(t)]\| \leq \alpha_{AB}$, and that 
    $$\max_{|s-t|\leq h }\norm{ [B(t), e^{\I A (s-t)} B(s) e^{-\I A (s-t)}]} \leq  C_{AB}h^{\theta}$$ for a non-negative real number $\theta$ and a constant $C_{AB}$ which might depend on $A$ and $B(t)$. 
    Then for any $0 < \epsilon < 1, T > \epsilon$, qHOP can implement an operation $W$ such that $\|W - U(T,0)\| \leq \epsilon$ with failure probability $\mathcal{O}(\epsilon)$ and the following cost: 
    \begin{enumerate}
        \item $\mathcal{O}\left( \left(\alpha_B T + \frac{C_{AB}^{1/(1+\theta)}T^{1+1/(1+\theta)}}{\epsilon^{1/(1+\theta)}} \log\left(\frac{C_{AB}T}{\epsilon}\right) \right)\log \left(\frac{(\alpha_{AB}+\beta_B)T}{C_{AB}\epsilon}\right) \right)$ uses of $O_A$, its inverse or controlled version, 
        \item $\mathcal{O}\left( \alpha_B T + \frac{C_{AB}^{1/(1+\theta)}T^{1+1/(1+\theta)}}{\epsilon^{1/(1+\theta)}} \log\left(\frac{C_{AB}T}{\epsilon}\right) \right)$ uses of $O_B(j)$, its inverse or controlled version, 
        \item $\mathcal{O}\left( \left(n_B + \log \left(\frac{(\alpha_{AB}+\beta_B)T}{C_{AB}\epsilon}\right)\right)\left( \alpha_B T + \frac{C_{AB}^{1/(1+\theta)}T^{1+1/(1+\theta)}}{\epsilon^{1/(1+\theta)}} \log\left(\frac{C_{AB}T}{\epsilon}\right) \right)\right)$ one- or two-qubit gates, 
        \item $\mathcal{O}(1)$ additional ancilla qubits.
    \end{enumerate}
\end{theorem}

\subsubsection{Scaling of the commutator}

Now we establish the bound for the commutator, \emph{i.e.} estimation of the parameters $C_{AB}$ and $\theta$. 
We again first study the most general case where we only assume that the norm of $A$ is much larger than the norm of $B(t)$. 
Error bound of this case can be directly obtained by applying \cref{thm:Mag1_interaction} and plugging in the concrete forms of the commutators. 

\begin{lemma}[Bounds for commutators in the interaction picture Hamiltonian simulation]\label{lem:interaction_commutator}
    For any $h > 0$, if $\max_{t\in[0,T]}\|B(t)\| \leq \alpha_B$, $\max_{t\in[0,T]}\|B'(t)\| \leq \beta_B$ and $\max_{t\in[0,T]}\|[A,B(t)]\| \leq \alpha_{AB}$, then we have 
        \item \begin{equation}
            \max_{|s-t|\leq h }\norm{ [B(t), e^{\I A (s-t)} B(s) e^{-\I A (s-t)}]} \leq \min\left\{ 2\alpha_B^2, 2\alpha_B(\alpha_{AB}+\beta_B)h \right\}. 
        \end{equation}
\end{lemma}
\begin{proof}
    It suffices to show that the commutator is bounded by both terms in the bracket on the right hand side. 
    The first bound follows trivially by 
    \begin{align} 
        &\quad \left\|[B(t),e^{\I A (s-t)}B(s)e^{-\I A (s-t)}]\right\| \nonumber \\
        &\leq \left\|B(t)e^{\I A (s-t)}B(s)e^{-\I A (s-t)}\right\| + \left\|e^{\I A (s-t)}B(s)e^{-\I A (s-t)}B(t)\right\| \nonumber\\
        &\leq 2\alpha_B^2. 
    \end{align}
    To prove the second bound, we use the fundamental theorem of calculus to obtain 
    \begin{equation}
        B(t) = B(s) + \int_{s}^{t} B'(\tau) d\tau. 
    \end{equation}
    Then 
    \begin{align}
        &\quad \left\|[B(t),e^{\I A (s-t)}B(s)e^{-\I A (s-t)}]\right\| \nonumber \\
        & \leq \left\|[B(s),e^{\I A (s-t)}B(s)e^{-\I A (s-t)}]\right\| + \left\|\left[\int_s^t B'(\tau)d\tau,e^{\I A (s-t)}B(s)e^{-\I A (s-t)}\right]\right\| \nonumber \\
        & \leq \left\|[B(s),e^{\I A (s-t)}B(s)e^{-\I A (s-t)}]\right\| + 2\alpha_B\beta_B h. 
    \end{align}
    To further bound the first term, we view $s$ as a fixed time, denote $t' = s-t$ and use the fundamental theorem of calculus with respect to  $t'$ to get 
    \begin{equation} \label{eqn:AB_fund_thm_calc}
        e^{\I A t'} B(s) e^{-\I A t'} = B(s) + \I \int_0^{t'} e^{\I A \tau} [A,B(s)] e^{-\I A \tau}  d\tau, 
    \end{equation}
    and thus 
    \begin{align}
        \left\|[B(t),e^{\I A (s-t)}B(s)e^{-\I A (s-t)}]\right\| &\leq \left\|\left[B(s),\I \int_0^{t'} e^{\I A \tau} [A,B(s)] e^{-\I A \tau}  d\tau\right]\right\| + 2\alpha_B\beta_B h \nonumber \\
        & \leq 2\alpha_B \alpha_{AB}h + 2\alpha_B\beta_B h.   \label{eqn:AB_bound_comm}
    \end{align}
\end{proof}

\cref{lem:interaction_commutator} implies two possible sets of parameters $(C_{AB}, \theta)$: $C_{AB} = 2 \alpha_B^2, \theta = 0$, or $C_{AB} = 2 \alpha_B (\alpha_{AB}+\beta_B), \theta = 1$. 
The following result can then be proved by plugging these two possible choices back into \cref{thm:Mag1_interaction}. 

\begin{cor}[Long-time complexity of qHOP in the interaction picture]
    For any $0 < \epsilon < 1, T > \epsilon$, if $\max_{t\in[0,T]}\|B(t)\| \leq \alpha_B$, $\max_{t\in[0,T]}\|B'(t)\| \leq \beta_B$ and $\max_{t\in[0,T]}\|[A,B(t)]\| \leq \alpha_{AB}$, then qHOP can implement an operation $W$ such that $\|W - U(T,0)\| \leq \epsilon$ with failure probability $\mathcal{O}(\epsilon)$ and 
    \begin{equation}
    \begin{split}
        &\mathcal{O}\Big( \min\left\{\frac{\alpha_B^2 T^{2}}{\epsilon}\log\left(\frac{\alpha_B T}{\epsilon}\right), \alpha_B T + \frac{\alpha_B^{1/2} (\alpha_{AB}+\beta_B)^{1/2} T^{3/2}} {\epsilon^{1/2}} \log\left(\frac{\alpha_B (\alpha_{AB}+\beta_B)T}{\epsilon}\right)  \right\}\\
        &\times  \log \left(\frac{(\alpha_{AB}+\beta_B)T}{\epsilon}\right) \Big)
    \end{split}
    \end{equation}
    uses of $O_A$ and $O_B(j)$. 
\end{cor}

Finally, for the special case when the original Hamiltonian $H = A+B$ with time-independent $B$, we can choose $\beta_B = 0$ and the complexity estimates can be slightly simplified as follows. 
\begin{cor}[Long-time complexity of qHOP in the interaction picture with time-independent B]
    For any $0 < \epsilon < 1, T > \epsilon$, if $\|B\| \leq \alpha_B$ and $\|[A,B]\| \leq \alpha_{AB}$, then qHOP can implement an operation $W$ such that $\|W - U(T,0)\| \leq \epsilon$ with failure probability $\mathcal{O}(\epsilon)$ and 
    \begin{equation}
    \begin{split}
        &\mathcal{O}\Big( \min\left\{\frac{\alpha_B^2 T^{2}}{\epsilon}\log\left(\frac{\alpha_B T}{\epsilon}\right), \alpha_B T + \frac{\alpha_B^{1/2} \alpha_{AB}^{1/2} T^{3/2}} {\epsilon^{1/2}} \log\left(\frac{\alpha_B \alpha_{AB}T}{\epsilon}\right)  \right\}\\
        &\times  \log \left(\frac{\alpha_{AB}T}{\epsilon}\right) \Big)
    \end{split}
    \end{equation}
    uses of $O_A$ and $\widetilde{O}_B$. 
\end{cor}

\subsection{Superconvergence for simulating the Schr\"odinger equation in the interaction picture}\label{sec:superconvergence}

Now we focus on the improved error bound for the commutator in the case of simulating the Schr\"odinger equation in the interaction picture, i.e. when $A =- \Delta$ and $B = V(x)$. Generically, the preconstant for the second order convergence is proportional to $\norm{[A,B]}$, which is $\Or(N)$ in the spatially discretized setting (see \cite[Appendix A]{AnFangLin2021}), and $N$ is the number of the spatial grid points. In the case of the Schr\"odinger equation, the commutator $[B, e^{\I As} B e^{-\I As}]$ provides further cancellation, and we have $C_{AB} = C_B ( = C_V)$ independent of $A$ and $\theta = 1$. In other words, qHOP exhibits \textit{superconvergence} for the Schr\"odinger equation simulation. This leads to a surprising second order convergence rate in the operator norm, and the preconstant is independent of $\norm{[A,B]}$. This significantly reduces the overhead cause by the a large  $N$, which can be in many cases the bottleneck for such real-space Hamiltonian simulation \cite{KivlichanWiebeBabbushEtAl2017,AnFangLin2021}. 

Recall the general cases of $A$ and $B$, though a naive application of the Taylor expansion
\begin{equation*}
    e^{-\I As} = I - \I A s - \frac{A^2}{2} s^2 + \cdots
\end{equation*}
or the fundamental theorem of calculus \cref{eqn:AB_fund_thm_calc} can provide a linear scaling in terms of $h$, the resulting bound depends on the norm of the commutator $\norm{[A,B]}$.
Another way to view this is that $\Delta$ is an unbounded operator, and one cannot directly perform the Taylor expansion for $e^{\I \Delta s}$ for any $s>0$.
Nevertheless, in the case of the Schr\"odinger equation simulation, the dependence of $A$ can be removed by leveraging the tools from pseudo-differential calculus (see e.g. \cite{Stein1993,zworski_book}).

For simplicity of the analysis, here we  consider the Schr\"odinger equation in the continuous space. Numerical experiments indicate that similar results can hold for the discretized version, as well as for other boundary conditions. On an  intuitive level, the idea is to keep the full information of the unitary $e^{-\I As}$ and rewrite it in a special way,
which makes use of the cancellation resulting from the oscillations. Different from the Taylor expansion and the fundamental theorem of calculus, the whole time-dependent part $e^{\I As} B e^{-\I As}$ can be written as a pseudo-differential operator, whose commutator with $B = V$ can be shown to be of order $h$ with scaling only dependent on $V$ and the dimension $d$.

We consider the Schwartz space ${\displaystyle {\mathcal {S}}(\mathbb {R} ^{m})}$ defined as the space of all smooth functions acting on $\mathbb {R} ^{m}$ that are rapidly decreasing at infinity along with all partial derivatives, and denote its dual space -- the space of tempered distributions -- as ${\displaystyle {\mathcal {S}'}(\mathbb {R} ^{m})}$ that consists of the continuous linear functional on ${\displaystyle {\mathcal {S}}(\mathbb {R} ^{m})}$~\cite{Stein1993,Stein2011}. Despite the technicality, the Schwartz functions and tempered distributions are natural objects to work with for the Fourier integral operators, because the Fourier transform is an automorphism of the Schwartz space and also of its dual space due to duality. Intuitively, one can think of a tempered distribution as a distribution (generalized function) that grows no faster than polynomials at infinity.
To prepare for the proof, we introduce the Weyl quantization \cite{zworski_book}, a type of pseudo-differential operator as a useful tool simplifying the calculations. Given $a(x, p) \in {\displaystyle {\mathcal {S}'}(\mathbb {R} ^{2d})}$, the Weyl quantization  $\Op(a)$ of the symbol $a(x,p)$ acting on $u \in {\displaystyle {\mathcal {S}}(\mathbb {R} ^{d})}$ is defined by the formula
\begin{equation}
    \Op(a) u(x) : = (2\pi)^{-d} \int_{\mathbb{R}^{2d}} a\left(\frac{x+y}{2}, p\right) e^{\I p\cdot (x-y)} u(y) \, dy \, dp.
\end{equation}
Such a quantization procedure in fact agrees well with physical intuitions. For example, the quantization of $x$ is the position operator $\hat x$ acting on $u$ as $\Op(x) u  = x u = \hat x u$, a multiplication operator and the quantization of $p$ is the momentum operator $\Op(p) = -\I \nabla = \hat p$. More generally, one has
\[
\Op(a(x)) u(x)= a(x) u(x), \quad \Op(p^\alpha) = (-\I \nabla)^\alpha u,
\]
for any multi-index $\alpha$. In particular, when $a(x, p)$ is a smooth function bounded together with all of its derivatives, $\Op(a)$ defines a bounded operator mapping from $L^2(\mathbb{R}^{d})$ to $L^2(\mathbb{R}^{d})$. To be specific, the operator norm of such a linear transformation $\mathcal{A}: L^2(\mathbb{R}^{d}) \to L^2(\mathbb{R}^{d})$ is defined as
\begin{equation*}
    \norm{\mathcal{A}}_{\mathcal{L}(L^2)} : = \inf_{u\in L^2(\mathbb{R}^d), u\ne 0}\frac{\norm{\mathcal{A} u}_{L^2}}{\norm{u}_{L^2}}.
\end{equation*}
Thanks to the Calder\'on-Vaillancourt theorem (see \cite[Theorem 4.23]{zworski_book} for Weyl quantization and \cite[Theorem 2.8.1]{Martinez2002} for more general pseudo-differential operators), $\Op(a)$ can be estimated as 
\begin{equation}\label{eqn:c-v-thm}
    \norm{\Op(a)}_{\mathcal{L}(L^2)} \leq C \sum_{|\alpha|\leq M d} \norm{\partial^\alpha a}_{L^\infty},
\end{equation}
for some constant $C$ and $M$, and the bound depends on a finite number of derivatives of $a$, growing linearly with the dimension $d$. Therefore, one can work with test functions $u \in L^2(\mathbb{R}^{d})$. Note that regularity assumption of $a(x,p)$ ensures that it belongs to the symbol class $S_{2d}(1)$. The definition of the quantization can be extended to other symbol classes (see, e.g., \cite[Chapter 4.4]{zworski_book} and \cite[Chapter 2]{Martinez2002}) and the $L^2$ boundedness may also be relaxed without assuming all derivatives bounded \cite{Cordes1975,Boulkhemair1999}. For simplicity, we work with symbols that are smooth functions bounded together with all of their derivatives.

\begin{lemma}[Bound for the commutator for the Schr\"odinger equation in the interaction picture]\label{lem:bound_commutator_realspace}
For a smooth function $V$ bounded together with all of its derivatives and $0 < h \leq 1$, we have
\begin{equation} 
  \max_{s \in [-h, h]} \norm{[V(x), e^{\I s \Delta} V(x) e^{-\I s \Delta}]}_{\mathcal{L}(L^2)} \leq C_V h,
\end{equation}
where $C_V$ is some constant depending only on $V$ and the dimension $d$.
\end{lemma}

\begin{proof} 

We divide the presentation of the proof into three steps. First, we calculate the commutator $[\Delta , \Op(a)]$ for any $a(x,p)$ smooth and bounded together with all of its derivatives, for any $u \in {\displaystyle {\mathcal {S}}(\mathbb {R} ^{d})}$ using integration by parts:
\begin{align*}
    & [\Delta, \Op(a)] u(x) 
    \\
    = & (2\pi)^{-d} \int_{\mathbb{R}^{2d}} \Delta_x \left( a\left(\frac{x+y}{2}, p\right) e^{\I p\cdot (x-y)} \right) u(y)\, dy \, dp \\
    & - (2\pi)^{-d} \int_{\mathbb{R}^{2d}}  a\left(\frac{x+y}{2}, p\right) e^{\I p\cdot (x-y)} \Delta_y u(y) \, dy \, dp
    \\
     = & (2\pi)^{-d} \int_{\mathbb{R}^{2d}} \Delta_x \left( a\left(\frac{x+y}{2}, p\right) e^{\I p\cdot (x-y)} \right) u(y)\, dy \, dp \\
    & - (2\pi)^{-d} \int_{\mathbb{R}^{2d}} \Delta_y\left( a\left(\frac{x+y}{2}, p\right) e^{\I p\cdot (x-y)} \right) u(y) \, dy \, dp. 
\end{align*}
A straightforward calculation reveals that  
\begin{align*}
    &\Delta_x \left( a\left(\frac{x+y}{2}, p\right) e^{\I p\cdot (x-y)} \right) \\
    = & \Delta_x a\left(\frac{x+y}{2}, p\right) e^{\I p\cdot (x-y)} + \I \nabla_x  a\left(\frac{x+y}{2}, p\right) \cdot p e^{\I p\cdot (x-y)} -  a\left(\frac{x+y}{2}, p\right) |p|^2 e^{\I p\cdot (x-y)},
    \\
    &\Delta_y \left( a\left(\frac{x+y}{2}, p\right) e^{\I p\cdot (x-y)} \right) \\
    = & \Delta_x a\left(\frac{x+y}{2}, p\right) e^{\I p\cdot (x-y)} - \I \nabla_x  a\left(\frac{x+y}{2}, p\right) \cdot p e^{\I p\cdot (x-y)} -  a\left(\frac{x+y}{2}, p\right) |p|^2 e^{\I p\cdot (x-y)}.
\end{align*}
This shows that 
\begin{equation} \label{eqn:lap_a_exact}
    [\Delta, \Op(a)]  = 2\I \Op(\nabla_x  a \cdot p).
\end{equation}
Therefore, one can then calculate the following difference
\begin{align*}
    & e^{\I s \Delta} V e^{-\I s \Delta} - \Op\left( V(x-2ps) \right)
    \\
    = & \int_0^s \frac{d}{d\tau}\left( e^{\I \tau \Delta} \Op\left( V(x-2p(s - \tau)) \right) e^{-\I \tau \Delta} \right)  \, d\tau 
    \\
    = & \int_0^s  e^{\I \tau \Delta} \left( \I [\Delta , \Op\left( V(x-2p(s- \tau)) \right)]
    + 2 \Op\left( p \cdot \nabla  V(x-2p(s- \tau)) \right) \right) e^{-\I \tau \Delta} \, d\tau = 0,
\end{align*}
where in the last line we used \cref{eqn:lap_a_exact} with the symbol $a$ chosen as $V(x-2p(s- \tau))$. Thanks to the assumption of $V$ and the Calder\'on-Vaillancourt theorem \cref{eqn:c-v-thm}, $\Op\left( V(x-2ps) \right)$ now defines a bounded operator mapping from $L^2(\mathbb{R}^{d})$ to $L^2(\mathbb{R}^{d})$.

The second step of the proof is to estimate the commutator $[V, e^{\I s \Delta} V e^{-\I s \Delta}]$, namely,
\begin{align*}
    [V, \Op\left( V(x-2ps) \right)] u(x) 
    = (2\pi)^{-d} \int_{\mathbb{R}^{2d}} \left(V(x) - V(y) \right) V\left(\frac{x+y}{2} - 2ps\right) e^{\I p\cdot (x-y)} u(y) \, dy \, dp,
\end{align*}
for all $u \in L^2(\mathbb{R}^d)$. Note that
\[
V(x) - V(y) = \int_{0}^{1} \frac{d}{d\tau}V(y + \tau (x-y))\, d\tau = \int_{0}^{1} (x-y) \cdot \nabla V(y + \tau (x-y))\, d\tau,
\]
together with integration by parts in $p$ we have
\begin{align*}
  &  [V, \Op \left(V (x-2ps) \right)] u(x)
  \\
    = & (2\pi)^{-d}  \int_{\mathbb{R}^{2d}}  \int_{0}^{1} (x-y) \cdot \nabla V(y + \tau (x-y)) V\left(\frac{x+y}{2} - 2ps\right) e^{\I p\cdot (x-y)} u(y) \, d\tau \, dy \, dp 
  \\
    = & (2\pi)^{-d} 2 s  \int_{\mathbb{R}^{2d}}  \int_{0}^{1}   \nabla V(y + \tau (x-y)) \cdot \nabla V\left(\frac{x+y}{2} - 2ps\right)  u(y)  e^{\I p\cdot (x-y)} \, d\tau \, dy \, dp. 
\end{align*}
Therefore, it suffices to show the $L^2$-norm of
\[
\Theta : =(2\pi)^{-d}  \int_{\mathbb{R}^{2d}}  \int_{0}^{1}   \nabla V(y + \tau (x-y)) \cdot \nabla V \left(\frac{x+y}{2} - 2ps\right)  u(y)  e^{\I p\cdot (x-y)} \, d\tau \, dy \, dp 
\]
is bounded by some constant, which is the third step of the proof. Note that $\Theta$ is a pseudo-differential operator of the symbol 
\[
b(x, y, p) : = \int_{0}^{1}   \nabla V(y + \tau (x-y)) \cdot \nabla V \left(\frac{x+y}{2} - 2ps\right) \, d\tau.
\]
acting on $u$, which can be written as
\[
\Theta = \widetilde{\Op}(b) u : = (2\pi)^{-d}  \int_{\mathbb{R}^{2d}}  b(x,y,p) u(y)  e^{\I p\cdot (x-y)} \, dy \, dp. 
\]
Thanks to the assumption on $V$, we can apply the Calder\'on-Vaillancourt theorem \cite[Theorem 2.8.1]{Martinez2002}
\[
\norm{\widetilde{\Op}(b)}_{\mathcal{L}(L^2)} \leq C \sum_{|\alpha|\leq M} \norm{\partial^\alpha b}_{L^\infty},
\]
where the constant $M$ depends only on the dimension, yielding the bound only depending on the dimension $d$ and $V$ together with its derivatives.
This completes the proof.
\end{proof}

Though the proof is based on continuous operators, we remark that this commutator error bound is also preserved in the discretized setting (see \cref{sec:num}), namely it is bounded by $C_B s$, where $C_B$ only depends on the matrix $B$. 
This yields an available choice of parameters $(C_{AB},\theta)$ specified in \cref{thm:Mag1_interaction} for simulating the Schr\"odinger equation, \emph{i.e. } $C_{AB} = C_B, \theta = 1$. 
Plugging such a choice back to \cref{thm:Mag1_interaction} gives the cost estimate of qHOP to simulate the Schr\"odinger equation as follows\footnote{Here we absorb constants related to $B$ or $V(x)$ to the $\Or$ notation since they are bounded due to the regularity of the function $V(x)$.}:  
    \begin{enumerate}
        \item $\mathcal{O}\left( \frac{T^{3/2}}{\epsilon^{1/2}} \log\left(\frac{T}{\epsilon}\right) \log \left(\frac{NT}{\epsilon}\right) \right)$ uses of $O_A$, its inverse or controlled version, 
        \item $\mathcal{O}\left( \frac{T^{3/2}}{\epsilon^{1/2}} \log\left(\frac{T}{\epsilon}\right) \right)$ uses of $O_B$, its inverse or controlled version, 
        \item $\mathcal{O}\left(  \frac{T^{3/2}}{\epsilon^{1/2}} \log\left(\frac{T}{\epsilon}\right) \left(n_B + \log \left(\frac{NT}{\epsilon}\right)\right)\right)$ one- or two-qubit gates, 
        \item $\mathcal{O}(1)$ additional ancilla qubits.
    \end{enumerate}

\section{Numerical results} \label{sec:num}
In this section, we demonstrate the numerical results for the Schr\"odinger equation simulation in the interaction picture. 
For simplicity, we consider the following Hamiltonian
\begin{equation}
H= -\Delta + V(x), \quad V(x) = \cos(4x), \quad x\in [-\pi,\pi]
\label{eqn:ham_lap_v}
\end{equation}
with periodic boundary conditions. Here $A$ corresponds to the discretized $-\Delta$ using a second order finite difference scheme, and $B$ the discretized $V(x)$, respectively.

\begin{figure}
    \centering
    \includegraphics[width=.55\textwidth]{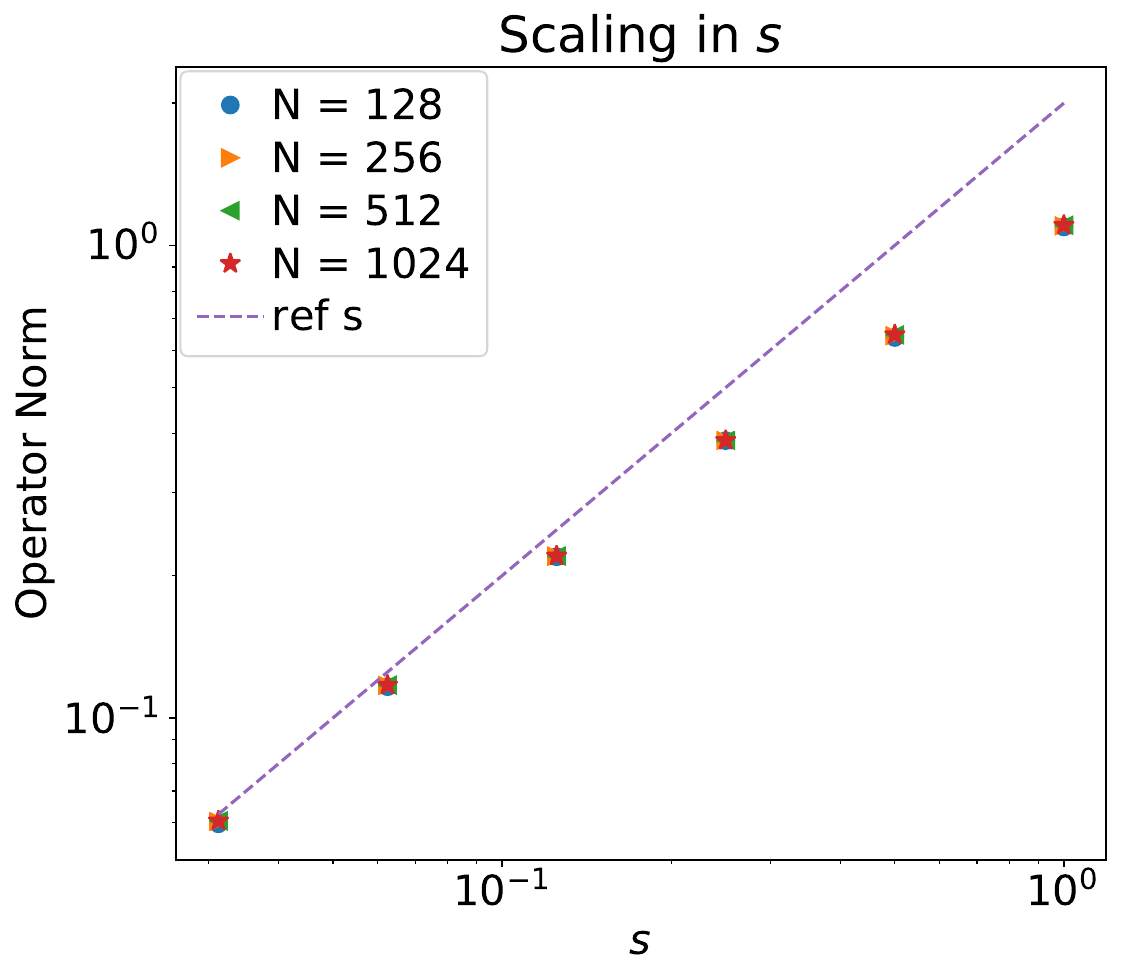}
    \caption{Log-log plot of the scaling of the norm of the commutator $[B, e^{\I A s} B e^{-\I A s}]$ in terms of $s$ for various $N$. Here $A$ and $B$ are the discrete Laplacian and potential operators, respectively. The reference line is for asymptotic scaling in $s$, and $N$ denotes the number of the grid points used in spatial discretization. }
    \label{fig:comm_scaling_s}
\end{figure}

First, we verify the statement in \cref{lem:bound_commutator_realspace} numerically for this periodic Hamiltonian. \cref{fig:comm_scaling_s} plots the norm of the commutator $[B, e^{\I A s} B e^{-\I A s}]$ in terms of the time $s$ for $N = 128, 256, 512, 1024$ in the log-log scale. 
It can be seen that the norm of this commutator grows at most linearly in $s$, and its preconstant is independent of $N$ as in \cref{lem:bound_commutator_realspace}.

\begin{figure}
    \centering
    \includegraphics[width=.55\textwidth]{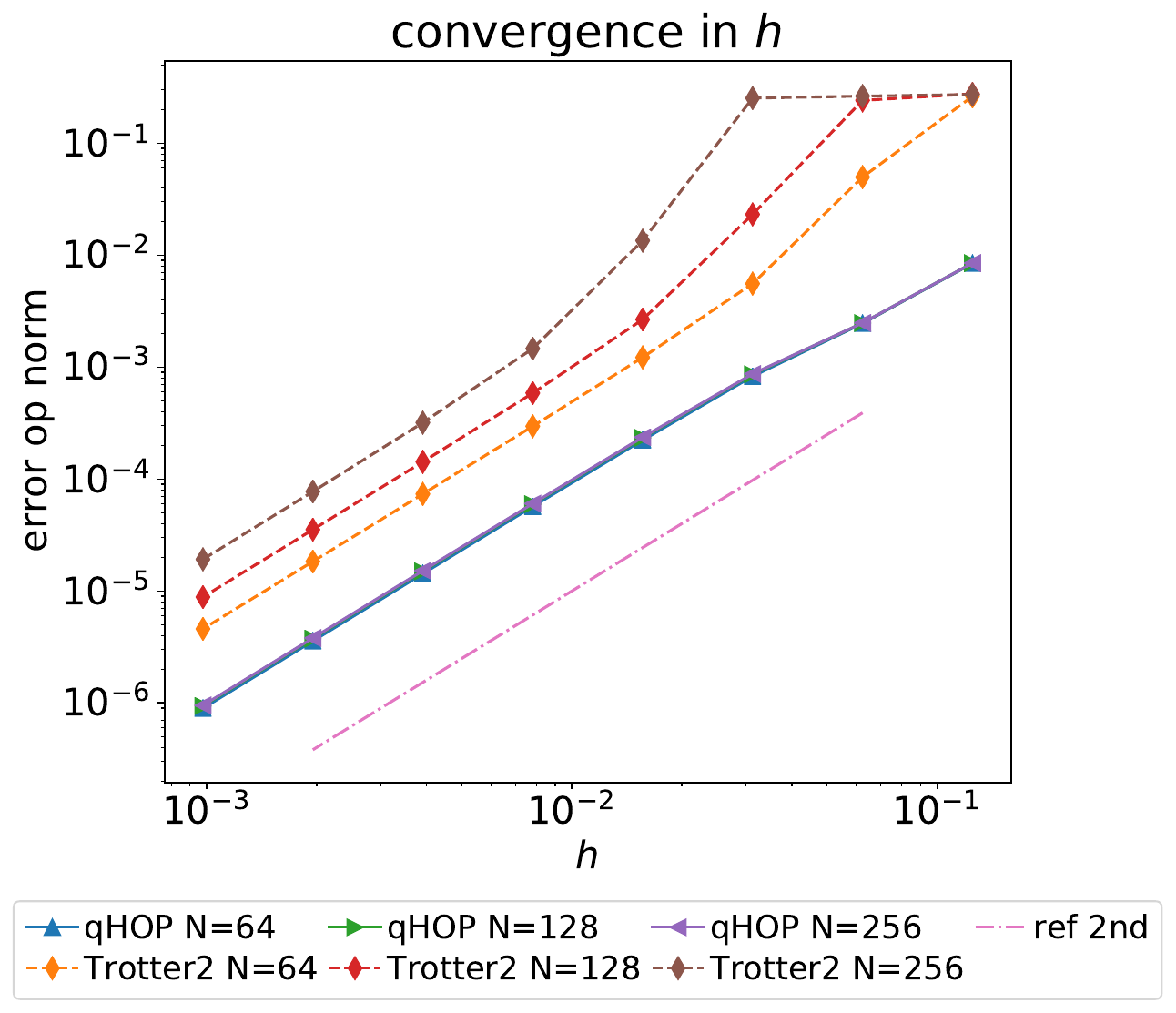}
    \caption{Log-log plot of the errors in the operator norm for various time step sizes $h$. The spatial discretization is finite difference. Both qHOP and the second-order Trotter formula exhibit second order convergence. However, the error of qHOP is smaller than that of the second-order Trotter formula, and does not grow as the number of the grid points in spatial discretization $N$ increases. The reference line demonstrates the asymptotic scaling.}
    \label{fig:lapV_conv_dt}
\end{figure}

We then present the convergence rate with respect to the time step $h$. \cref{fig:lapV_conv_dt} plots the operator norm error versus  $h$ in the log-log scale. Here the values of $h$ are chosen as $2^{-3}, 2^{-4}, \cdots, 2^{-10}$ and $M$ is fixed to be a sufficiently large number $2^{24}h$, so that the quadrature error becomes negligible. The system is simulated until the final time $t = 0.5$ and the number of spatial discretization $N$ is fixed as $128$.
It can be seen that both qHOP and the second-order Trotter formula exhibit the second order convergence in $h$. However, the error of qHOP is an order of magnitude smaller than that of the second-order Trotter formula, in particular, the Trotter error grows with respect to the increase of $N$ while qHOP remains the same.

\begin{figure}
    \centering
    \includegraphics[width=.55\textwidth]{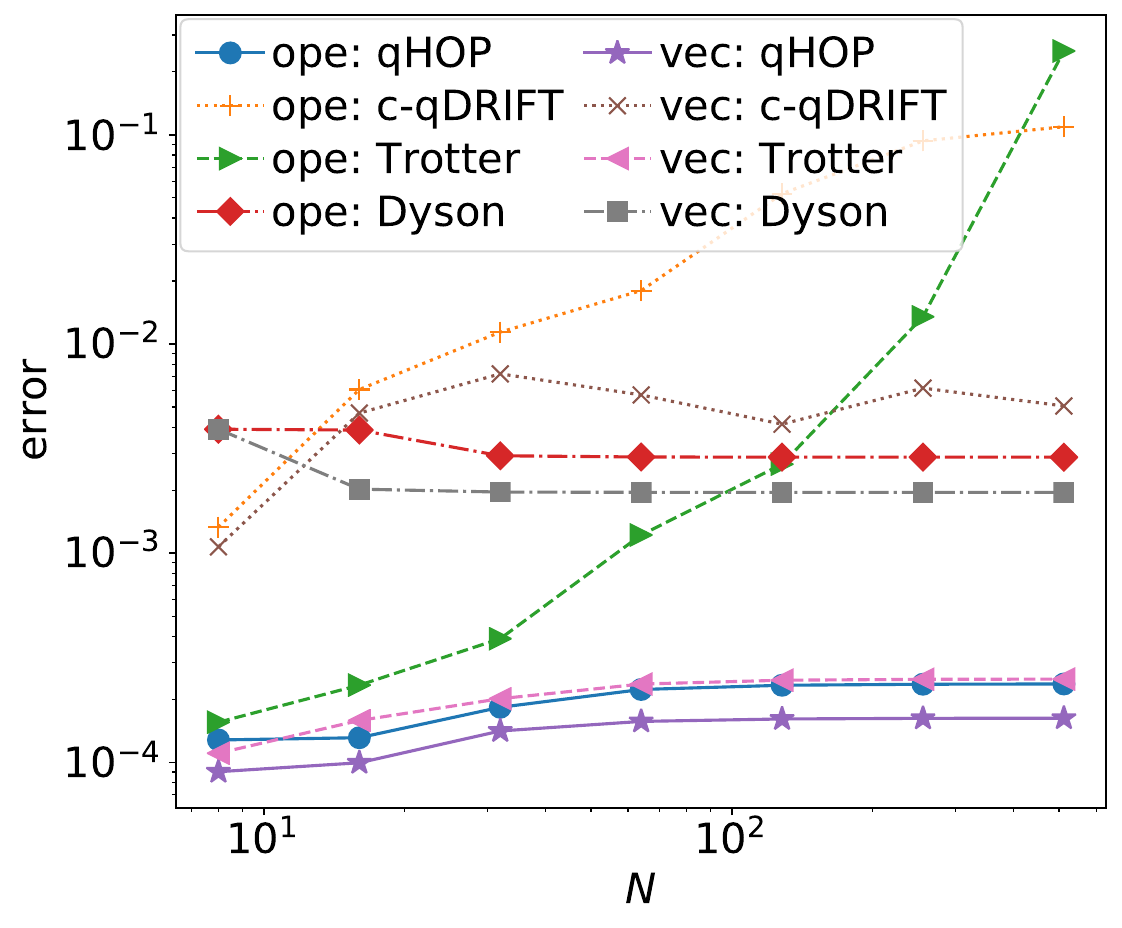}
    \caption{Log-log plot of the errors in both the operator norm and the vector norm for various numbers of grid points used in spatial discretization denoted by $N$. The spatial discretization is finite difference. The error in operator norm is labeled as ``ope'' while the one in vector norm as ``vec''. The vector norm is computed using a low-frequency Gaussian wavepacket \cref{eqn:gau_wvpkt_freq1} as the initial wavefunction. The number of quadrature points used in qHOP and that of the truncated Dyson series are the same.}
    \label{fig:lapV_N}
\end{figure}

The third numerical example compares the scaling of the vector norms and the operator norms, respectively. To evaluate the vector norm error, we take the initial vector $\vec{v}$ as the discretization of a low-frequency Gaussian wavepacket
\begin{equation} \label{eqn:gau_wvpkt_freq1}
    \exp(-4(x+1)^2) \exp(\I (x+1)). 
\end{equation}
The time step size $h$ is fixed to be $1/64$. The number of spatial grids $N$ are chosen as $8$, $16$, $32$, $64$, $128$, $256$ and $512$. We remark that the first order numerical quadrature requires a potentially large number of quadrature points. This can significantly increase the cost of the classical computation, but only introduces a logarithmic factor to the cost of the quantum simulation. In the numerics of this and the next example, we use the second order trapezoidal quadrature rule to approximate the integral \cref{eqn:int_quadrature} instead with $512$ quadrature points. Note that this is comparable to take $M$ to be around $2^{18}$ for a first order quadrature implementation. We compare the performance of qHOP, continuous qDRIFT (c-qDRIFT), the second-order Trotter formula (Trotter2) and first-order truncated Dyson (Dyson1) up to a final time $t = 0.5$. The number of quadrature points used in truncated Dyson series is the same as that in qHOP. We measure the one-instance error for the continuous qDRIFT method, where the probability distribution to be sampled from is a uniform distribution (this is because the norm $\norm{H_I(s)} = \norm{B}$ is a constant). The errors for both the operator and vector norms are plotted in \cref{fig:lapV_N}. 
We find that qHOP exhibits the smallest error (sometimes orders of magnitude smaller) measured both in the operator norm and in the vector norm.
In terms of the operator norm, the errors of both qHOP and Dyson1 does not grow with respect to $N$, while the error of Trotter2 grows with respect to $N$. Furthermore, the error of qHOP is much smaller than that of Dyson1, because the latter is a first order scheme while qHOP is of second order. As for the vector norm, with respect to a low-frequency initial condition, Trotter2 does not grow with respect to $N$ which agrees with the result shown in \cite{AnFangLin2021}.

\begin{figure}
    \centering
    \includegraphics[width=.55\textwidth]{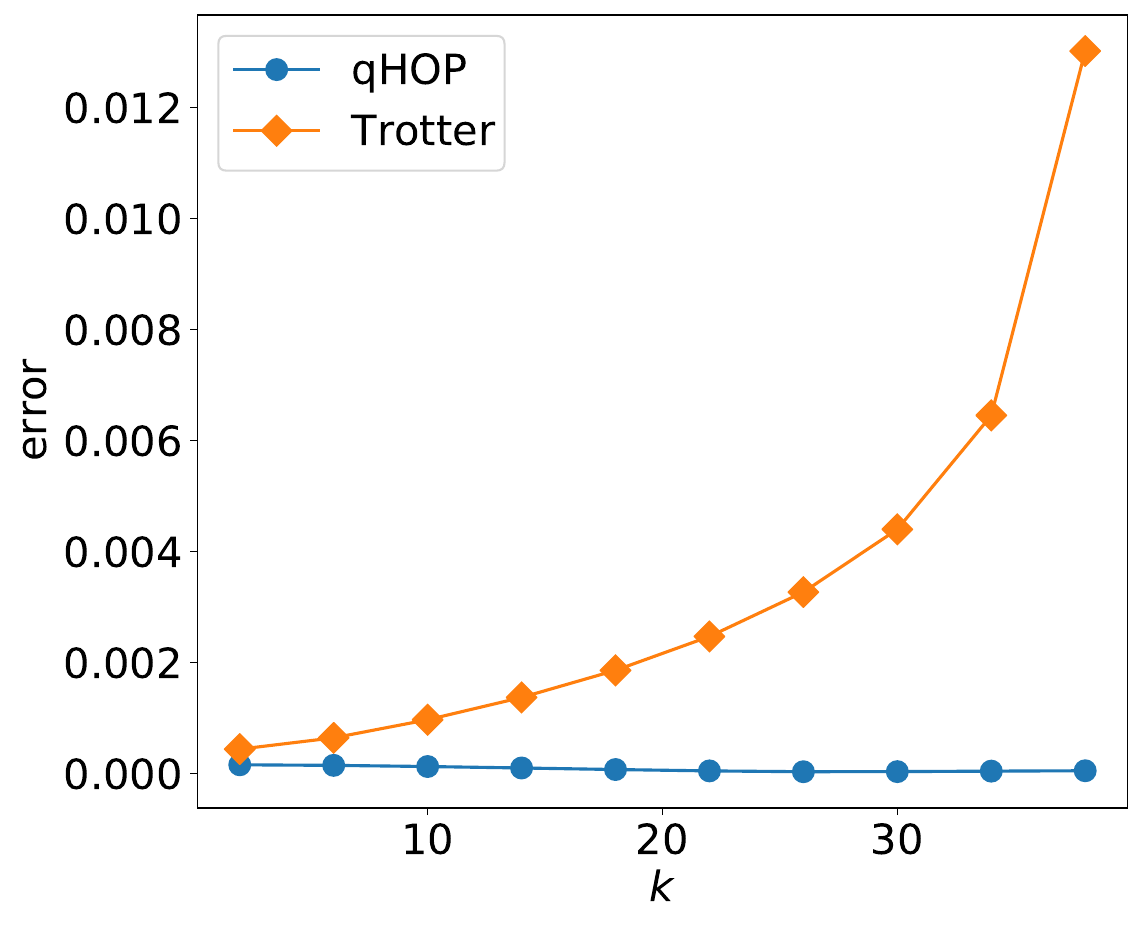}
    \caption{The errors in the vector norm for the initial wavepacket \cref{eqn:gau_wvpkt_freqk} with various frequency $k$. The spatial discretization is finite difference. It can be seen that qHOP outperforms Trotter2 in the vector norm scaling.}
    \label{fig:lapV_k}
\end{figure}

In the last example, we demonstrate more carefully the performance of both qHOP and Trotter2 when measuring the vector norm errors. Note that $N$-independent vector norm error bounds can be achieved (see \cite{AnFangLin2021}), but the preconstant of the error bound can still depend on the smoothness of the initial condition. On the other hand, qHOP has $N$-independent operator norm error bounds, and hence its vector norm error can be upper bounded by the operator norm error independent of the smoothness of the initial vector. To illustrate this point, we consider a series of initial vectors obtained by discretizing the Gaussian wavepacket with various frequencies $k$ given as
\begin{equation} \label{eqn:gau_wvpkt_freqk}
    \exp(- 20 (x+1)^2) \exp(\I k (x+1)).
\end{equation}
The time step size $h$ is fixed to be $1/64$ and the grid number $N$ is fixed to be $512$. The system is simulated till the final time $t = 0.5$. It can be seen in \cref{fig:lapV_k} that as the frequency $k$ of the initial vector increases, the vector norm error of qHOP does not grow while that of Trotter2 increases significantly.

\section{Conclusion and discussion}
In this work, we proposed the quantum highly oscillatory protocol (qHOP), a simple quantum algorithm for the time-dependent Hamiltonian simulation that provides a unified solution to treat the fast variation and/or the large operator norm of a time-dependent Hamiltonian. The resulting algorithm explores the commutator scaling like the Trotter-type algorithm, exhibits $L^1$-norm scaling like the continuous qDRIFT method, and remains insensitive to the rapid change of $H(t)$ as the truncated Dyson series. The construction of the method can be interpreted as a first order truncation of the Magnus series, followed by a quantum numerical quadrature implemented via the linear combination of unitary technique.
In this case, the prepare oracle is simply a set of Hadamard gates. This converts the time-dependent Hamiltonian simulation into a series of time-independent Hamiltonian simulation problems, which can in turn be efficiently performed using techniques such as the quantum singular value transformation (QSVT).

As an application, qHOP provides an alternative approach to simulate in the interaction picture. Though we mainly focus on the scenario of $H(t) = A + B(t)$, the extension of our qHOP formulation to the more general case $H(t) = a(t) A + B(t)$ is also possible. Assuming that $A$ is fast-forwardable and $a(t)$ is a scalar function whose antiderivative can be accurately computed with negligible extra cost, the interaction Hamiltonian can be taken as $$H_I(t) = e^{\I A \int_0^t a(s)\, ds} B(t) e^{-\I A \int_0^t a(s)\, ds}.$$ 
Plugging in \cref{eqn:Mag1_IP_u1_quadrature} yields the desired formulation, and the quantum circuit of implementing the block encoding of the linear combination of interaction picture Hamiltonians needs to be changed accordingly, which we do not detail here.
The advantage is particularly prominent for simulating the Schr\"odinger equation, where qHOP exhibits superconvergence and achieves a second order convergence rate.
The query complexity is only logarithmic in the number of grids $N$. 
Numerical results indicate that qHOP can be much more accurate than the first order truncated Dyson method and the continuous qDRIFT method, and can outperform the Trotter methods both in operator norms, as well as in vector norms with oscillatory initial vectors.

Similar to the widely used second order Trotter formula, we think that qHOP provides a suitable balance between efficiency and accuracy, and can be a useful quantum algorithm for practical treatment of highly oscillatory problems in a wide range of scenarios. 
Although the $L^1$-norm scaling of qHOP we establish in this work is in terms of the time average of Hamiltonian spectral norm, it may be possible to obtain an improved error bound that depends on the average of the commutators. 
This is because the one-step error of qHOP indeed follows a local $L^1$ scaling in terms of the commutators, and it would be\ interesting to study whether qHOP equipped with the $L^1$-norm scaling in commutators can provide further speedup for certain physical systems.
We also remark that the high order generalization of qHOP is also possible, by means of truncating the Magnus series to higher orders.
This however inevitably re-introduces the time-ordering operator, and its implementation requires quantum control logic.  
In certain scenarios, the efforts may be compensated by the higher order of accuracy.
Our preliminary numerical results indicate that by truncating the Magnus series to second order and by adopting a sufficiently accurate numerical quadrature, the resulting method can again exhibit superconvergence for simulating the Schr\"odinger equation, and achieves a \textit{fourth order} convergence rate.
Following this strategy to high orders, we expect that the cost of the resulting method can be insensitive to $\norm{H'(t)}$, exhibit commutator scalings, and depend poly-logarithmically on the precision parameter $\epsilon$.

\section*{Acknowledgments:} 

This work was partially supported by the NSF Quantum Leap Challenge Institute (QLCI) program through grant number OMA-2016245 (D.F.), by the NSF under Grant No. DMS-1652330 (D.A.), and by Department of Energy under Grant No. DE-SC0017867  and the Quantum Systems Accelerator program  (L.L.). 
D.A. acknowledges the support by the Department of Defense through the Hartree Postdoctoral Fellowship at QuICS. 
L.L. is a Simons Investigator.

\bibliographystyle{abbrvurl}
\bibliography{ref_bib_doi}

\clearpage
\appendix

\section{Existing algorithms for time-dependent Hamiltonian simulation}

We summarize several existing quantum algorithms for time-dependent Hamiltonian simulation and briefly discuss their scalings. 
For completeness, we reintroduce the notations that will be used in showing the scalings of existing algorithms. We consider both the general time-dependent Hamiltonian simulation problem and the simulation in the interaction picture with a specific focus on the Schr\"odinger equation.

For the general problem, we consider 
\begin{equation}
    \I \partial_t \ket{\psi(t)} = H(t) \ket{\psi(t)}, \quad 0 \leq t \leq T,
\end{equation}
where $H(t)$ is a time-dependent Hamiltonian. 
The exact evolution operator is given as 
\begin{equation}
    U(T,0) = \mathcal{T}e^{-\I \int_0^T H(s)ds}
\end{equation}
where $\mathcal{T}$ is the time-ordering operator. 
Our goal is to construct another unitary operator $U_{\text{num}}(T,0)$ such that $\|U_{\text{num}}(T,0)-U(T,0)\| \leq \epsilon$ where $\|\cdot\|$ denotes the spectral norm. 
Typical approaches include dividing the entire interval $[0,T]$ into $L$ equi-distant segments and approximating the exact operator by local numerical propagator, and we will use $h = T/L$ to denote the time step size whenever applicable. 
The same as those in \cref{tab:main_result_1}, we assume $\max_{s\in[0,T]} \|H(s)\| \leq \alpha$, $T^{-1}\int_0^T \|H(s)\| ds \leq \overline{\alpha}$, $\max_{s,t\in[0,T]}\|[H(s),H(t)]\| \leq \widetilde{\alpha}^2$, and $\max_{s,t\in[0,T]}\|[H'(s),H(t)]\| \leq \widetilde{\beta}$.

The second model we consider is 
\begin{equation}
    \I \partial_t \ket{\psi(t)} = (A+B(t)) \ket{\psi(t)}, \quad 0 \leq t \leq T
\end{equation}
where $A$ possibly has a large spectral norm but can be fast-forwarded, and $B(t)$ is bounded. 
The same as those in \cref{tab:main_result_2}, we assume $\|A\| \leq \alpha_A$ is large but $e^{-\I A t}$ can be fast-forwarded, $\max_{t\in[0,T]}\|B(t)\| \leq \alpha_B$, $\max_{t\in[0,T]}\|B'(t)\| \leq \beta_B$, and $\max_{t\in[0,T]}\|[A,B(t)]\| \leq \alpha_{AB}$. 
In the interaction picture, we transform the state by $\ket{\psi_I(t)} = e^{\I A t} \ket{\psi(t)}$ and solve 
\begin{equation}
    \I \partial_t \ket{\psi_I(t)} = H_I(t) \ket{\psi_I(t)},
\end{equation}
where $H_I(t) = e^{\I A t}B(t)e^{-\I A t}$. 
In the case of the Schr\"odinger equation, $A$ is the discretized Laplacian operator $(-\Delta)$ and $B(t)$ is the discretized potential operator, with $N$ denoting the number of basis functions (grid points) used in spatial discretization.

\subsection{Trotter and generalized Trotter methods}

We discuss the generalized Trotter formulae for the time-dependent Hamiltonian simulation, and the (standard) Trotter formulae for the time-independent case. In order to apply Trotter-type algorithms, the Hamiltonian needs to be in the form of $H(t) = \sum_{j =1}^S H_j(t)$. For simplicity, we consider the case of two terms. $H(t) = H_1(t) + H_2(t)$. The generalized Trotter formulae \cite{HuyghebaertDeRaedt1990} is given by
\begin{equation} \label{eqn:gen_trotter}
\mathcal{T}e^{-\I \int_t^{t+h}H(s) \,ds}
\approx \mathcal{T} e^{-\I \int_t^{t+ h } H_1(s) \,ds} \mathcal{T} e^{-\I \int_t^{t+ h} H_2(s) \,ds},
\end{equation}
and \cite[Eq (2.3)]{HuyghebaertDeRaedt1990} provides the error bound for this short time evolution
\[
\int_t^{t+ h} \,ds \int_t^s \, d  u \norm{ [H_1(s), H_2(u)] }  \leq \frac{1}{2}  \max_{s,u \in [t, t+h]} \norm{ [H_1(s), H_2(u)] }  h^2.
\]
Hence the long time error for the evolution till time $T$ can be estimated as
\[
\frac{1}{2}  \max_{s,u \in [t, t+h]} \norm{ [H_1(s), H_2(u)] } \frac{T^2}{L}
\]
where $L$ is the number of the equi-length segments dividing the time interval $[0,T]$.
However, it is worth pointing out that the generalized Trotter formulae \cref{eqn:gen_trotter} by itself is not an algorithm, in the sense that some further treatment -- such as qHOP, continuous qDRIFT and truncated Dyson series -- is still required to implement the time-ordering operators.
One exception is that $H(t)$ is of the form of a controlled Hamiltonian 
\[
H(t) = f_1(t) H_1 + f_2(t) H_2,
\]
where $f_1(t)$ and $f_2(t)$ are some control functions, in which case the resulting unitaries are free of the time ordering operator. 

For the time-independent Hamiltonian $H = A + B$, time-independent Trotter-type formulae~\cite{ChildsSuTranEtAl2020} can be directly applied. 
In particular, the first order Trotter formula is 
\[
e^{-\I (A+B) h} \approx e^{-\I B h} e^{-\I A h}. 
\]
The one-step local Trotter error bound given by \cite[Proposition 15]{ChildsSuTranEtAl2020} reads
\[
\frac{h^2}{2}\norm{[A,B]},
\]
and thus the global Trotter error is bounded by 
\[
\frac{\norm{[A,B]}}{2} \frac{T^2}{L}. 
\]
Therefore, to achieve the $\epsilon$-approximation of the unitary evolution in the operator norm, the query complexity for the first order Trotter formula is \cite[Corollary 12]{ChildsSuTranEtAl2020}
\[
\Or\left(\frac{\norm{[A,B]}T^2}{\epsilon}\right) 
\]
and in the case of the real-space Hamiltonian simulation becomes 
\[
\Or\left( \frac{N T^2}{\epsilon}\right),
\]
because $\norm{[A,B]} = \Or(N)$ \cite[Appendix A]{AnFangLin2021}. 

For the second order Trotter formula 
\[
e^{-\I (A+B) h} \approx e^{-\I A h/2} e^{-\I B h} e^{-\I A h/2}, 
\]
the cost to achieve an $\epsilon$-approximation can be estimated in a similar way, which gives the query complexity for the second order Trotter formula as \cite[Corollary 12]{ChildsSuTranEtAl2020}
\[
\Or\left(\frac{\left( \norm{[B,[B,A]]} +  \norm{[A,[A,B]]} \right)^{1/2} T^{3/2}}{\epsilon^{1/2}}\right) 
\]
and in the case of the real-space Hamiltonian simulation becomes 
\[
\Or\left( \frac{N T^{3/2}}{\epsilon^{1/2}}\right),
\]
because $\norm{[B,[B,A]]} = \Or(N)$ and $ \norm{[A,[A,B]]}  = \Or(N^2)$ \cite[Appendix A]{AnFangLin2021}.

\subsection{Monte Carlo method} 

The idea of the Monte Carlo method proposed in~\cite{PoulinQarrySommaEtAl2011} is to approximate the exact evolution operator first by first-order generalized Trotter and then to approximate the resulting integral by Monte Carlo sampling. The goal of this algorithm is to get rid of the dependence on the time-derivatives of $H(t)$.

For a single Hamiltonian $H(t)$, the local evolution operator can be approximated as 
\begin{equation}
\begin{split}
    \mathcal{T}e^{-\I \int_{t}^{t+h} H(s) ds} &\approx e^{-\I \int_{t}^{t+h} H(s) ds}  \\
    &\approx e^{-\I \frac{h}{m}\sum_{j=1}^m H(s_j)} \\
    &\approx \prod_{j=1}^m e^{-\I \frac{h}{m} H(s_j)}, 
\end{split}
\end{equation}
where $s_j$'s are random variables sampled uniformly in $[t,t+h]$.  
According to the proof of our \cref{lem:Mag1_discretization_error}, the approximation error of the first approximation is bounded by \footnote{We remark that here we use our improved error estimate for the first step of approximation. In~\cite{PoulinQarrySommaEtAl2011} this step is directly bounded by $\mathcal{O}(h^2 \max_{t}\|H(t)\|^2)$. } $$\mathcal{O}(h^2\max_{s,\tau\in[t,t+h]}\|[H(\tau),H(s)]\|). $$
According to~\cite{PoulinQarrySommaEtAl2011}, the error due to the second step of the Monte Carlo approximation is bounded by $$h\max_{s\in[t,t+h]}\|H(s)\|/\sqrt{m},$$ and the error of the third step is the standard Trotter error bounded by $$\mathcal{O}( h^2 \max_{s,\tau\in[t,t+h]}\|[H(\tau),H(s)]\|). $$
Summarizing all these together, we can bound the local error by 
\begin{equation}
    \mathcal{O}( h^2 \max_{s,\tau\in[t,t+h]}\|[H(\tau),H(s)]\| + h\max_{s\in[t,t+h]}\|H(s)\|/\sqrt{m}).
\end{equation}
According to \cref{lem:general_commutator}, we can further bound the local error by 
\begin{equation}
    \mathcal{O}( h^2 \min\left\{\widetilde{\alpha}^2, \widetilde{\beta} h\right\} + h\alpha /\sqrt{m}), 
\end{equation}
and the corresponding global error by 
\begin{equation}
    \mathcal{O}\left(  \min\left\{\widetilde{\alpha}^2T^2/L,\widetilde{\beta}T^3/L^2 \right\} + \alpha T/\sqrt{m}\right). 
\end{equation}
To bound the error by $\epsilon$, it suffices to choose 
\begin{equation}
    L = \mathcal{O}\left( \min\left\{\frac{\widetilde{\alpha}^2 T^2}{\epsilon}, \frac{\widetilde{\beta}^{1/2} T^{3/2}}{\epsilon^{1/2}}\right\}\right), \quad m = \mathcal{O}\left(\frac{\alpha^2 T^2}{\epsilon^2}\right),
\end{equation}
and the total number of queries to $e^{-i t H(s)}$ is $\mathcal{O}(Lm)$.  Note that the number of quadrature points $m$ contributes a  multiplicative factor  to the query complexity.

Now we consider the complexity of the Monte Carlo method in the interaction picture for general $A$ and $B(t)$. 
The under the interaction picture, $H_I(t) = e^{\I A t}B(t)e^{-\I A t}$ and thus the local error can be bounded by 
\begin{equation}
    \mathcal{O}( h^2 \max_{s,\tau\in[t,t+h]}\|[H_I(\tau),H_I(s)]\| + h\max_{s\in[t,t+h]}\|H_I(s)\|/\sqrt{m}). 
\end{equation}
According to \cref{lem:interaction_commutator}, we can further bound the local error by 
\begin{equation}
    \mathcal{O}( h^2 \min\left\{\alpha_B^2,\alpha_B(\alpha_{AB}+\beta_B)h\right\} + h\alpha_B /\sqrt{m}),
\end{equation}
and the corresponding global error by 
\begin{equation}
    \mathcal{O}(  \min\left\{T^2\alpha_B^2/L, T^3\alpha_B(\alpha_{AB}+\beta_B)/L^2\right\} + T\alpha_B /\sqrt{m}). 
\end{equation}
To bound the error by $\epsilon$, it suffices to choose 
\begin{equation}
    L = \mathcal{O}\left(\min\left\{\frac{T^2\alpha_B^2}{\epsilon}, \frac{T^{3/2}\alpha_B^{1/2}(\alpha_{AB}+\beta_B)^{1/2}}{\epsilon^{1/2}}\right\}\right), \quad m = \mathcal{O}\left(\frac{T^2\alpha_B^2}{\epsilon^2}\right),
\end{equation}
and the total number of queries to $e^{-\I A t}$ and the input model of $B(t)$ is $\mathcal{O}(Lm)$. 

Finally, in the case of Schr\"odinger equation where $B(t)$ is assumed to be smoothly bounded, we can use \cref{lem:bound_commutator_realspace} to directly bound the local error by 
\begin{equation}
    \mathcal{O}\left(h^3 + h/\sqrt{m}\right),
\end{equation}
which gives the global error bound as 
\begin{equation}
    \mathcal{O}\left(T^3/L^2 + T/\sqrt{m}\right). 
\end{equation}
Therefore it suffices to choose 
\begin{equation}
    L = \mathcal{O}\left(\frac{T^{3/2}}{\epsilon^{1/2}}\right), \quad m = \mathcal{O}\left(\frac{T^2}{\epsilon^2}\right). 
\end{equation}

\subsection{Continuous qDRIFT} 
The idea of the continuous qDRIFT is to approximate the exact quantum channel by certain stochastic protocol. Assume the spectral norm $\norm{H(\tau)}$ is known \textit{apriori} or can be accurately upper bounded, the algorithm approximates the ideal quantum channel corresponding to the exact evolution $U(t,0)$ defined as
\begin{align*}
    \mathcal{E}(t,0)(\rho)=U(t,0)\rho U^\dagger(t,0) 
    = \mathcal{T} \exp \biggl(-i\int_{0}^{t}\mathrm{d}\tau\, H(\tau)\biggr)
    \rho
    \mathcal{T} \exp \dagger\biggl(-i\int_{0}^{t}\mathrm{d}\tau\, H(\tau)\biggr).
\end{align*}
by a mixed unitary channel given by
\begin{equation*} 
\mathcal{E}(t,0)(\rho) \approx \mathcal{U}(t,0)(\rho)=\int_{0}^{t}\mathrm{d}\tau\, p(\tau)e^{-i \frac{H(\tau)}{p(\tau)}}\rho e^{i \frac{H(\tau)}{p(\tau)}},
\end{equation*}
where $p(\tau)$ is a probability density function defined for $0\leq \tau\leq t$,
\[
p(\tau):=\frac{\norm{H(\tau)}}{\int_0^t \norm{H(\tau)} \,d\tau}.
\]
The quantum channel $\mathcal{U}(t,0)(\rho)$ is then implemented via a classical sampling protocol: for any input state $\rho$, one randomly sample $\tau$ from the distribution $p(\tau)$ and perform $e^{-\I H(\tau)/p(\tau)}$. \cite[Theorem 7]{BerryChildsSuEtAl2020} shows that one can divide the interval $[0,T]$ into $0 = t_0 < t_1 < \cdots < t_L = T$ such that when the continuous qDRIFT protocol is performed on each sub-interval, the long time simulation error in the diamond norm for the quantum channels is bounded by 
\[
\frac{4 \left( \int_0^T \norm{H(\tau)} \,d\tau \right)^2}{L}.
\]
To obtain an $\epsilon$-approximation of the ideal quantum channel using continuous qDRIFT protocol, the query complexity is 
\[
\Or\left( \frac{ \left( \int_0^T \norm{H(\tau)} \, d \tau \right)^2}{\epsilon} \right),
\]
where $n$ is the number of qubits that $H$ acts on,
assuming the probability distribution $p(\tau)$
can be efficiently sampled. Therefore, the query complexity in $T$ and $\epsilon$ is given as  
\[
\Or\left( \frac{\left( \frac{1}{T} \int_0^T \norm{H(\tau)} \, d \tau \right)^2 T^2}{\epsilon} \right) = \Or \left(\frac{\overline{\alpha}^2 T^2}{\epsilon}\right).
\]

Now we consider the complexity of the continuous qDRIFT in the interaction picture for general $A$ and $B(t)$. 
Straightforward calculations show that 
\begin{equation} \label{eqn:L1_norm_IP}
\frac{1}{T} \int_0^T \norm{H_I(\tau)} \, d \tau =  \frac{1}{T} \int_0^T \norm{B(t)} \, d \tau \leq \max_{t\in[0,T]}\norm{B(t)},
\end{equation}
which gives the query complexity 
$\Or \left( \alpha_B^2 T^2 /\epsilon \right)$.
In the case of the real-Hamiltonian simulation, $\norm{B} = \Or(1)$ and hence the query complexity becomes $\Or(T^2/\epsilon)$. Note that \cref{eqn:L1_norm_IP} also shows that for the time-independent Hamiltonian simulation in the interaction picture, the $L^1$ norm scaling is the same as the maximum norm scaling $T \max_{\tau \in [0, T]} \norm{H(\tau)}$.

\subsection{Truncated Dyson series method}
Truncated Dyson series method utilizes the Dyson series
\begin{align*}
    U(t,0) 
    &= I - \I \int_0^t H(t_1) dt_1 - \int_{0}^t\int_0^{t_2}  H(t_2)H(t_1) dt_1 dt_2 + \cdots \\
    & = \sum_{k= 0}^\infty \frac{(-\I)^k}{k!}\int_0^t d t_1 \int_0^t d t_2 \cdots \int_0^t d t_k \mathcal{T} \left[H(t_1) H(t_2) \cdots H(t_k) \right].
\end{align*}
The complexity of higher order truncated Dyson series method has been carefully analyzed in~\cite{LowWiebe2019} for both general simulation and the Hamiltonian simulation in the interaction picture. 
In particular, to simulate the dynamics on $[0,T]$ within $\epsilon$ error for a time-dependent Hamiltonian satisfying $\max_t\|H(t)\| \leq \alpha$, query complexity to the $\text{HAM-T}_j$ input model is given as (\cite[Corollary 4]{LowWiebe2019})
\begin{equation*}
    \Or \left(\alpha T \log(\alpha T/\epsilon)\right). 
\end{equation*}
As we have mentioned in the introduction, truncated Dyson series method beyond first order 
contains time-ordering and hence requires clocking done by quantum control logic.
Therefore, instead of adaptively selecting the truncated order according to the error level, we will restrict ourselves on the first order truncated Dyson series method. 

The only difference in the complexity analysis of the first order method from the higher order method is the choice of the segments used to divide the time interval $[0,T]$. 
Specifically, we start with the Dyson series expansion on a short time interval $[0,h]$ that 
\begin{equation*}
    U(h,0) = I - \I \int_0^h H(t_1) dt_1 - \int_{0}^h\int_0^{t_2}  H(t_2)H(t_1) dt_1 dt_2 + \cdots. 
\end{equation*}
The approximation error by truncating at the first order can be bounded as 
\begin{equation*}
    \left\|U(h,0) - \left(I - \I \int_0^h H(t_1) dt_1\right)\right\| \leq \Or (\alpha^2 h^2). 
\end{equation*}
Following the proof of~\cite[Theorem 3]{LowWiebe2019}, for any $\alpha h \leq 1/2$, using the construction in \cref{eqn:Mag1_LCU}, we can implement a circuit with $\Or(1)$ query to HAM-T such that it is an $\Or(\alpha^2h^2 + h^2\max_{s\in[0,T]}\|H'(s)\|/M)$ approximation of $U(h,0)$. 
Notice that  the second part of the error is due to the approximation of the integral, and $M$ is the number of quadrature nodes. 
By choosing $M = \mathcal{O}\left(\max_{s\in[0,T]}\|H'(s)\|/\alpha^2\right)$, the local approximation error can then be bounded by $\Or(\alpha^2h^2)$. 
Finally, following the proof of~\cite[Corollary 4]{LowWiebe2019}, the approximation of the long-time evolution operator $U(T,0)$ can be constructed with error $\Or(L\alpha^2h^2)$, where $L$ is the number of the segments dividing $[0,T]$. 
Plugging $h = T/L$ into the error estimate, the global approximation error can then be bounded by $\Or(\alpha^2T^2/L)$. 
Therefore, in order to bound the error by $\epsilon$, it suffices to choose $L = \Or(\alpha^2T^2/\epsilon)$, which leads to $$\Or\left(\frac{\alpha^2T^2}{\epsilon}\right)$$
queries to $\text{HAM-T}_j$. 

The cost of applying truncated Dyson series method to the interaction picture example can be analyzed similarly, by noticing that the corresponding HAM-T$_j$ oracle can be implemented using $\mathcal{O}(\log(M))$ queries to $O_A$ and $\mathcal{O}(1)$ query to $O_B$, and that $M = \mathcal{O}((\alpha_{AB}+\beta_B)/\alpha_B^2)$, which leads to 
$$\Or\left(\frac{\alpha^2T^2\log((\alpha_{AB}+\beta_B)/\alpha_B)}{\epsilon}\right)$$
queries to $O_A$.

\end{document}